# A simple modification for improving inference of non-linear dynamical systems

Wan Yang[*], Jeffrey Shaman

Department of Environmental Health Sciences, Mailman School of Public Health, Columbia University, New York, New York

## Abstract

Particle and ensemble filters are increasingly utilized for inference, optimization, and forecast; however, both filtering methods use discrete distributions to simulate continuous state space, a drawback that can lead to degraded performance for non-linear dynamical systems. Here we propose a simple modification, applicable to both particle and ensemble filters, that compensates for this problem. The method randomly replaces one or more model variables or parameters within a fraction of simulated trajectories at each filtering cycle. This modification, termed space re-probing, expands the state space covered by the filter through the introduction of outlying trajectories. We apply the space re-probing modification to three particle filters and three ensemble filters, and use these modified filters to model and forecast influenza epidemics. For both filter types, the space re-probing improves simulation of influenza epidemic curves and the prediction of influenza outbreak peak timing. Further, as fewer particles are needed for the particle filters, the proposed modification reduces the computational cost of these filters.

## 1. Introduction

Data assimilation, or filtering, methods are commonly used in conjunction with state-space models and observations to perform system optimization, parameter estimation, and prediction. In application, the model system of interest is propagated forward through time to generate a prediction (i.e., a background, or a prior); whenever an observation becomes available the filter is then used to calibrate or update the prediction (i.e., making an analysis, or posterior). Through time the model-simulated system is optimized by these repeated prediction-update cycles. Two major families of data assimilation methods are often used—particle filters [1-3] and ensemble filters [4-6]. Both methods apply Monte Carlo approaches to initiate multiple system replicas (termed particles or ensemble members) that simulate the dynamics of a particular system; however, the methods differ in their implementation. Particle filters evaluate the likelihood of each particle given observations. Trajectories close to the observations are given greater weight and enhanced, while those farther away are given less weight and may eventually be eliminated during resampling [1-3]. In contrast, ensemble filters, e.g. the ensemble Kalman filter (EnKF), usually compute the update (i.e. posterior) of the observed variable based on the prior, the observation, and their respective estimated variances [4,6]. The unobserved model variables and model parameters are then adjusted based on their covariance with the observed variable. In theory, for both approaches, the updated state converges to trajectories best representing true system dynamics.

---
[*] *Corresponding author address*: Wan Yang, Department of Environmental Health Sciences, Mailman School of Public Health, Columbia University, 722 West 168th Street, 11th Floor, New York, NY 10032, Email: wy2202@columbia.edu

Both assimilation methods have inherent limitations. In particular, both use discrete distributions to represent continuous space [3,7] and over the course of repeated prediction-update cycles can become trapped in a small sub-region of the overall state space. This trapping is potentially problematic as it limits the ability of the model-filter framework to respond to shifts in the dynamics of the system. For example, model-filter frameworks can be used to estimate key epidemiological parameters associated with infectious disease outbreaks; however, should the characteristics of the culprit pathogen evolve over the course of an outbreak, a trapped system may be ill-equipped to handle this change in system dynamics. In the case of particle filters, the particles, as guided by observations, may first concentrate in a small region of state space; should system behavior shift, the particles may not span enough of state space to capture this new behavior. This shortcoming would result in particle impoverishment and an overall degradation of filter performance (Figure 1A and Figure S1A and S1B left panels). For ensemble filters, as the ensemble members are adjusted towards the observation, the ensemble variance decreases; should the ensemble spread over-narrow the filter can place too much confidence in the prior and diverge from observations (Figure S2A and S2B left panels).

Multiple strategies have been developed to circumvent these issues, including regularization for the particle filters [1,2] and inflation for the ensemble filters [8,9]. These corrections are generally applied equally to all trajectories; therefore, only a small adjustment can be made so as not to disturb the filtering process. These corrections in effect only expand the range of state space around the original region of convergence. As a result, the filter is less capable of promptly adapting in response to a dramatic shift in system behavior.

Here, we propose a simple modification, termed space re-probing (SR), that is capable of quickly adapting the system while not disturbing the filtering process. The idea is to modify certain key model variables or parameters for a small portion of trajectories immediately following each update. Those modified trajectories can then give rise to new, distinct paths that span a broader range of state space. *That is, rather than eliminating outlying trajectories, this strategy intentionally creates a few paths that probe* *the entirety of state space.* As illustrated in Figure 1B for a particle filter, in most instances those modified trajectories are outliers, and eliminated at the next assimilation checkpoint. However, should the system change unexpectedly, those outlier trajectories may span the shifted state space. Under such circumstances, they become decisively helpful; the filter can quickly magnify those close to the new unexpected observation and shift other trajectories towards the altered sub-state space.

For the ensemble filters, the mechanism is similar in that it creates a broader ensemble variance, which may not only span new unexpected observations (Figure S2A and S2B right panels) but also reduces confidence in the prior, thus placing greater confidence in the observation and preventing system divergence. In addition, the modified model variable or parameter alters the covariance between model variables and parameters, a feedback that may further accelerate the adaptation to changing system behaviors (Figure S2A and S2B right panels). .

The SR modification thus broadly expands the state space span of the particles/ensemble such that the filtering framework, be it a particle or ensemble filter, can more reliably contain trajectories capable of representing changing system dynamics. To illustrate this modification further, we present an application to the modeling and forecast of influenza epidemics.

## 2. Application example: modeling and forecast of influenza epidemics

Influenza epidemics occur annually during winter in temperate regions and continue to produce high morbidity, mortality, and economic burden [10-12]. Recent studies indicate that reliable prediction of outbreak peak timing and other epidemic characteristics is achievable and that these predictions could enable better prevention and preparedness for the disease [13,14]. Our previous work built six model-filter frameworks [15], with three particle filters —a particle filter with conditional resampling and regularization (PF)[1], maximum likelihood estimation via iterated filtering (MIF) [16,17], and particle Markov chain Monte Carlo (pMCMC) [18,19]—and three ensemble filters—the EnKF [6], ensemble adjustment Kalman filter (EAKF) [4,13,20], and rank histogram filter (RHF) [5]. Each filter was used in conjunction with the same susceptible-infected-removed-susceptible (SIRS) model [13] and weekly measures of influenza incidence to simulate and forecast influenza epidemics. All six model-filter frameworks are able represent and predict unimodal influenza outbreaks with good accuracy, but are less capable representing and predicting multi-modal outbreaks [15].

Current circulating human influenza viruses include 2 subtypes of influenza A (i.e., H1N1 and H3N2), as well as influenza B. For each A subtype and the B type, multiple strains, or variants, exist [21]; during a single influenza season, the co-circulation of multiple strains/subtypes often leads to multiple epidemic waves [22]. Consequently, should the dominant circulating strain/subtype change mid-season, the characteristics of an outbreak may shift as well. Within a state-space model of influenza, this shift will manifest as a change in the model variables and parameters, which represent the population dynamics of the host and biological characteristics of the virus. For example, during a flu season, lumped population susceptibility could increase when a new strain becomes dominant, rather than decrease as would happen had no shift occurred. In addition, key epidemiological parameters, such as the infectious period and transmissibility, could differ among strains/subtypes. Because our SIRS model does not include multiple strains, it is not able to explicitly simulate the co-circulation of multiple strains. The particle filters, without replenishment of particles, tend to be trapped in the sub-state space representing the past influenza strain and thus fail to re-tailor the system to the new strain. The ensemble filters, while in theory able to partially compensate for the model bias as the ensemble can migrate to new state space, may in practice still diverge if too much confidence is placed in the prior.

In light of these features, we first apply the SR modification to the susceptibility variable, $S$, in our SIRS model. The same modification is applied to all six aforementioned model-filter frameworks. Preliminary analysis was used to determine when (at every updating cycle vs. triggered only by certain events), how (random vs. selective replacement), how many (the percentage of trajectories), and with what values to replace $S$ during the filtering process; ultimately, we settled on random replacement of 2% of trajectories at every updating cycle, with values drawn from a uniform distribution bounded by 50-60% of population for seasonal epidemics (Supplemental Material). Modifications to other model parameters such as the infectious period, $D$, and the maximum reproductive number, $R_{0max}$, or a combination of these three variable/parameters, are also tested; these modifications yielded similar improvements (Supplemental Material). We only report the modification to $S$ in the main text.

## 3. Results

### 3.1 Retrospective fitting

All six aforementioned model-filter frameworks were run with the SR modification and used the same SIRS model and estimates of influenza incidence to simulate epidemic curves for the 2003-04 through 2012-13 seasons for 115 cities in the U.S. We calculated the root mean squared (RMS) error for each run and compared this metric with those produced by unmodified versions of each filtering framework. The SR modification significantly improves model representation of observed epidemic curves for all six filters (i.e., lower RMS error, 1-sided t-test, $p<2.2e-16$ for all 6 filters). Of 4740 runs (for each filter), 83.44%, 89.24%, 88.82%, 94.73%, 95.15%, and 74.57% had lower RMS error than their respective unmodified filter simulations (Figure 2). Figure 3 shows how the SR-modified filters perform for different cities and seasons. A few pinkish spots, indicating degraded performance, are present; however, in general, the modified filters perform comparably well or do marginally better for most cities (white to lightly bluish part in Figure 3). Improvement is more profound for the MIF and pMCMC during these seasons.

The SR modification also appears to improve parameter estimation, as it partially compensates for model bias. For example, by explicitly replacing low susceptible levels when strain dominance shifts, the modified filters may compensate for model bias in the non-strain specific SIRS. As a result, parameter estimates made with the SR modified filters, particularly the ensemble filters, tend to be more in line with levels estimated through alternate methods [23,24]. For instance, to match with some observational time series records, the unmodified EAKF yielded estimates of D>10 days. Such overestimates were less common for the modified filters (Figure 4).

## 3.2 Forecasts

We then used the SR-modified filters to generate forecasts of influenza outbreak peak timing (i.e., the week with highest influenza incidence) during the 2003-04 through 2012-13 epidemic seasons for 115 U.S. cities. Each forecast is generated following a training period of 3-28 weeks, in which past observations of influenza incidence are assimilated up to the time of prediction through repeated weekly prediction-update cycles using one of the six filters. The forecast itself then uses the optimized model variables and parameters inferred from the training as initial conditions and integrates the SIRS model forward to the end of each season [13,25]. Forecasts are generated for each week and city of each season. In general, the forecasts with the SR modification produce more accurate predictions; in particular, the predicted influenza incidence time series were able to match the observations better (lower RMS error, Fig. S11-S16). Here, we report the SR modification improvements for prediction of outbreak peak timing.

Summed over all cities and seasons, the SR filter predictions outperform their corresponding unmodified versions (Figure 5 last row). This improvement is generally evident for each specific season as well (Figure 5 and Table 1). None of the seasons had degraded performance using the SR-modified filters (paired t-test, Table 1). As some seasons experienced a 'clean' unimodal outbreak (e.g., 2003-04) and others experienced a multimodal outbreak (e.g., 2005-06), this outcome suggests that for seasons without abrupt shifts (unimodal outbreaks) the SR modification does not degrade the filtering process. Further, for seasons with multimodal outbreaks (e.g., 2005-06 and 2010-11), the SR filters profoundly improve peak timing prediction, especially for the MIF and pMCMC (Figure 5).

The SR modification benefits the two filter types in slightly different fashion. For the particle filters, the improvement increases when the epidemic unfolds later in the season (Figure 6). This finding is not unexpected. As unique particles deplete over time (due to elimination of those with low weight), the particle filters tend to suffer more severe particle impoverishment later in the season. The SR modification, which replaces susceptibility levels for a small portion of particles, replenishes the filtering framework and expands the exploration of state space (Figure S1A). This replenishment is particularly important during later stages of the inference process. However, it should be noted that, for the pMCMC, forecasts made very early in the season and at certain weeks approaching the peak of most seasons are less accurate with the SR modification (Figure 5 and Figure 6C). The unmodified pMCMC outperforms its modified version, as well as the other 5 filters for forecasts made 5-2 weeks before the observed peak [15]. For the ensemble filters, the magnitude and timing of improvement is less obvious (Figure 6 D-F). In addition, the SR modification seems to benefit some cities more than others (darker pinkish colors in certain rows, e.g., Arizona cities in the bottom rows, for most filters).

## 4. Discussion

Filtering methods are excellent tools for improving the simulation, inference, and prediction of dynamical systems. This study proposes a simple modification to further improve the performance of these methods. The strategy is to replace certain key variable(s)/parameter(s) for a small portion of the ensemble. Using the simulation and prediction of influenza epidemics as a specific example, we have shown that the SR modification is able to improve representation of historical influenza activity and increase the accuracy of epidemic peak timing forecasts. The modification is effective for both particle and ensemble filters.

The SR modification is based on a general mechanism that could work for other systems, though no strict mathematical proof is provided. The rationale is that key model variables/parameters can significantly impact the evolution of system dynamics; explicit modification of these variables/parameters for a small fraction of trajectories effectively replenishes an ensemble that may be trapped within a portion of state space. Therefore, while specific tuning is needed for each system of interest (e.g., determination of which variable/parameter to modify, the proportion of trajectories, and with what values), we anticipate that improvements similar to those shown in this study could be achieved.

The SR modification works by simply expanding the state space explored during the filtering process (Figure 1); such expansion is achieved at nominal computational cost, i.e., inclusion of an additional step for the SR modification. Further, in the case of the particle filters, fewer particles are required. For example, in this study, the SR particle filters were run with 3000 particles compared to 10,000 particles for the unmodified versions. This reduced particle number reduces the computational demand of the particle filters by around half.

Methods such as regularization for the particle filters and inflation for the ensemble filters have been applied to ease particle impoverishment and filter divergence. These two methods modify the SIRS model state (for the PF and ensemble filters all model variables and parameters; for the MIF and pMCMC only the model variables) of all trajectories around their original paths [15]. In contrast, the SR method alters only a small fraction of trajectories and probes a wider span of state space without disturbing the filtering process. Although only a specific model variable/parameter is modified, it is able to induce systematic changes to other model variables and parameters that further increase the diversity among the simulated trajectories (Fig S7 and S8). In this study, for the purpose of comparison, the SR filters were run in addition to regularization for the particle filters and inflation for the ensemble filters, as in the unmodified versions. Further tests applying only the SR modification without regularization/inflation may confirm whether the SR alone is sufficient to prevent particle impoverishment or filter divergence. If so, the SR modification may further lower the computational cost of filtering.

More rigorous testing is needed to confirm the improvements shown here. Particularly, future work should test how the SR modification performs in real-time forecasting, as opposed to the retrospective forecasts made in this study. Work should also be expanded to the simulation and forecast of pandemic outbreaks to determine if the modification might enable earlier detection of pandemic signals. Validation of its effectiveness for other applications is also warranted.

## 5. Material and Methods

This study builds on previous work [15], which compared the same six filters tested here, i.e., three particle filters (PF, MIF, and pMCMC) and three ensemble filters (EnKF, EAKF, and RHF), when applied to a susceptible-infected-removed-susceptible (SIRS) model [13] and historical ILI+ data (i.e., a metric of influenza activity, computed as municipal scale Google Flu Trends [26] estimates of weekly influenza like illness (ILI) multiplied by the weekly census division regional proportion of laboratory confirmed influenza viral positive samples from among a portion of clinically diagnosed ILI cases [22] [25]. ILI+ records collected from 2003-04 through 2012-13 epidemic seasons (i.e., 9 seasons excluding the 2009 pandemic) were used in this study. The SIRS model is as described in [13].

The unmodified versions of the six model-filter frameworks are as described in [15]. In this study, however, we calculated the simulated peak as the week with the highest ILI+ for the mean predicted ILI+ time series (i.e., the mean trajectory of all particles/ensemble). For the SR modified filters, an additional step, which replaces the susceptible level, $S$, for 2% of the particles/ensemble members, is included at the end of each prediction-update cycle, for each filter (see pseudocode in Supplemental Material). All versions of the ensemble filters were run with 300 ensemble members. The particle filters were run with different numbers of particles to achieve best performance; specifically, 10,000 particles were used for all three unmodified filters (consistent with [15]), while 3000 particles were used for the SR filters as 10,000 particles did not substantially improve SR filter performance (Figure S5, Supplemental Material).

In conjunction with the SIRS model, the SR filters were used to simulate historical ILI+ time series for each of the aforementioned 9 seasons for up to 115 U.S. cities (the number of cities used each year ranges from 66-115 and depends on the availability of ILI data), and used to retrospectively forecast outbreak peak timing. Results from the SR filters were compared with those from the unmodified versions.

**Table 1.** Change in the accuracy (Δacc) of influenza peak timing forecasts for each season. Δacc was calculated as the mean difference in forecast accuracy, averaged over all cities for each week, between forecasts produced by the SR-modified and the corresponding unmodified filter; numbers in the parentheses are 95% confidence intervals. $p$ values were calculated from 2-sided paired t-test. All those with $p<0.05$ (*) had a Δacc greater than 0, i.e., significantly higher accuracy by the SR-modified filters.

| | PF | | MIF | | pMCMC | | EnKF | | EAKF | | RHF | |
|---|---|---|---|---|---|---|---|---|---|---|---|---|
| | Δacc | p | Δacc | p | Δacc | p | Δacc | p | Δacc | p | Δacc | p |
| 2003-04 | 0.012 (0.0052, 0.020) | 0.0015 * | 0.041 (0.027, 0.054) | 2.3e-06 * | 0.011 (0.004, 0.019) | 0.0038 * | 0.0017 (-0.004, 0.0075) | 0.53 | 0.0024 (-0.0025, 0.0074) | 0.32 | 0.0089 (6e-04, 0.017) | 0.037 * |
| 2004-05 | 0.00042 (-0.0036, 0.0044) | 0.83 | 0.041 (0.014, 0.068) | 0.0045 * | -0.0021 (-0.025, 0.021) | 0.85 | -0.0046 (-0.0095, 0.00035) | 0.067 | 0.0059 (-0.0023, 0.014) | 0.15 | 0.00059 (-0.0067, 0.0079) | 0.87 |
| 2005-06 | 0.019 (0.0098, 0.027) | 0.00019 * | 0.076 (0.041, 0.11) | 0.00012 * | 0.029 (0.00031, 0.058) | 0.048 * | 0.0071 (-0.0013, 0.016) | 0.095 | 0.011 (0.004, 0.018) | 0.0033 * | 0.0017 (-0.0033, 0.0067) | 0.50 |
| 2006-07 | 0.0091 (-0.0021, 0.02) | 0.11 | 0.028 (-0.0044, 0.060) | 0.087 | -0.0044 (-0.037, 0.028) | 0.78 | -0.00054 (-0.0062, 0.0051) | 0.84 | 0.0077 (-1.8e-05, 0.015) | 0.050 | 0.0049 (-0.0018, 0.012) | 0.14 |
| 2007-08 | -0.0058 (-0.013, 0.0012) | 0.10 | 0.015 (0.0049, 0.025) | 0.0056 * | -0.011 (-0.022, 0.00092) | 0.070 | 0.0011 (-0.0072, 0.0094) | 0.79 | 0.0049 (-0.0037, 0.013) | 0.25 | 0.0054 (-0.00096, 0.012) | 0.093 |
| 2008-09 | -0.0053 (-0.016, 0.0055) | 0.32 | 0.0082 (-0.0054, 0.022) | 0.22 | -0.008 (-0.026, 0.0096) | 0.36 | -0.0033 (-0.0099, 0.0032) | 0.30 | -0.0045 (-0.016, 0.0065) | 0.40 | 0.0045 (-0.0017, 0.011) | 0.15 |
| 2010-11 | 0.01 (0.00083, 0.020) | 0.034 * | 0.11 (0.053, 0.16) | 0.00041 * | 0.053 (0.0082, 0.097) | 0.022 * | 0.027 (0.0095, 0.045) | 0.0041 * | 0.04 (0.017, 0.063) | 0.0015 * | -0.0012 (-0.0067, 0.0043) | 0.65 |
| 2011-12 | -0.0019 (-0.0058, 0.0021) | 0.34 | 0.0084 (-0.0031, 0.020) | 0.15 | -0.011 (-0.028, 0.0063) | 0.20 | -0.0045 (-0.015, 0.0058) | 0.38 | 0.0022 (-0.013, 0.018) | 0.77 | -0.00087 (-0.013, 0.011) | 0.88 |
| 2012-13 | 0.072 (0.032, 0.11) | 0.0012 * | 0.17 (0.088, 0.26) | 0.00038 * | 0.082 (0.028, 0.14) | 0.0046 * | 0.022 (0.01, 0.034) | 0.00092 * | 0.028 (0.013, 0.043) | 0.00095 * | 0.028 (0.015, 0.042) | 0.00025 * |

**Figure 1.** Mechanism for the SR modification in a particle filter. In the unmodified filter (A), particles are selected towards a narrower and narrower distribution consistent with past observations made during the filtering process (e.g., from t=0 to 20). If the system changes abruptly, this narrowed distribution may no long span the newly observed true state (e.g., at t=30). For the modified filter (B), at each filtering step, a few particles (in blue) are modified such that they may migrate away from the main stream. At each assimilation checkpoint, the weight of each reprobing particle is decreased if it is an outlier or increased if it better represents the new state. This illustration is adapted from van Leeuwen 2009 [2].

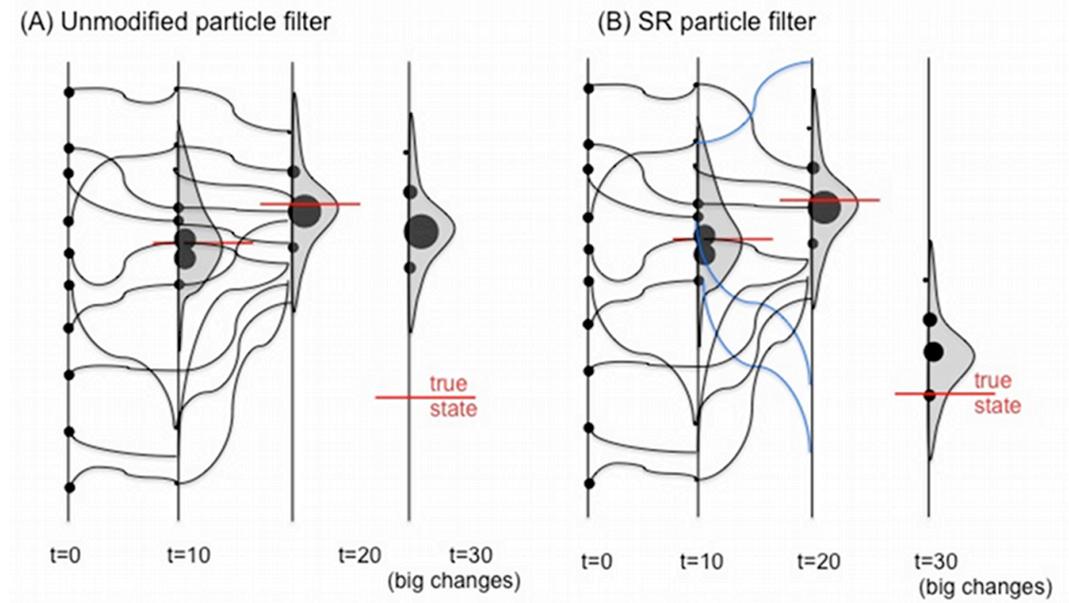

**Figure 2.** Distribution of the difference in root mean squared error (ΔRMS) when using the SR modified filter vs. the unmodified filter. ΔRMS is calculated as the averaged RMS error between the modeled and historical ILI+ time series (over 5 repeated runs) from the SR modified filter minus that from the unmodified filter; a negative ΔRMS indicates improved fit. The distribution of ΔRMS is shown for each filter (A-F).

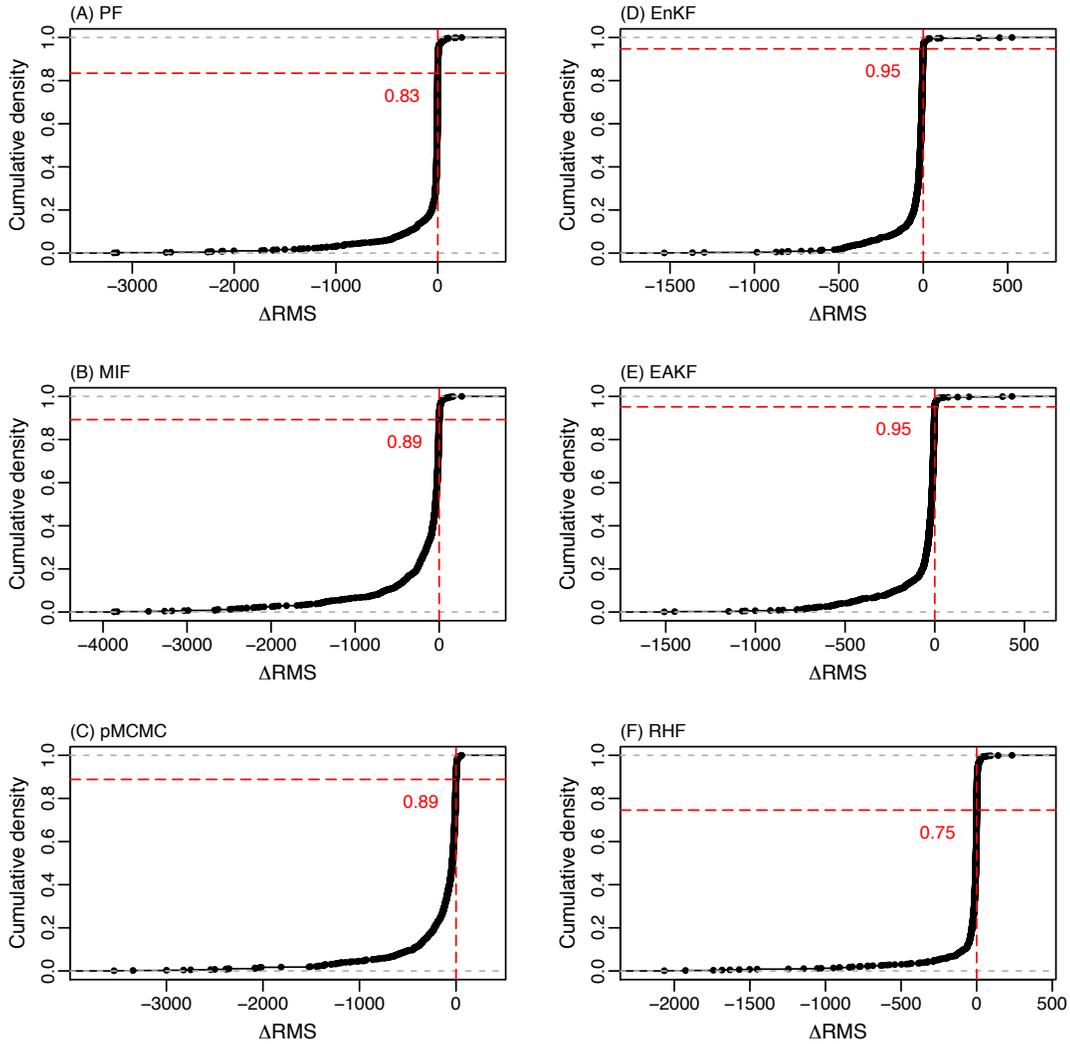

**Figure 3.** Difference in root mean squared error (ΔRMS) between the modeled and historical ILI+ time series for the 115 cities from 2003-04 (A) through 2012-13 (I) epidemic seasons. ΔRMS is calculated as the averaged RMS error (over 5 repeated runs) from the SR modified filter minus that from the unmodified filter. Grey cells are cities for which no observed data were available in a given year.

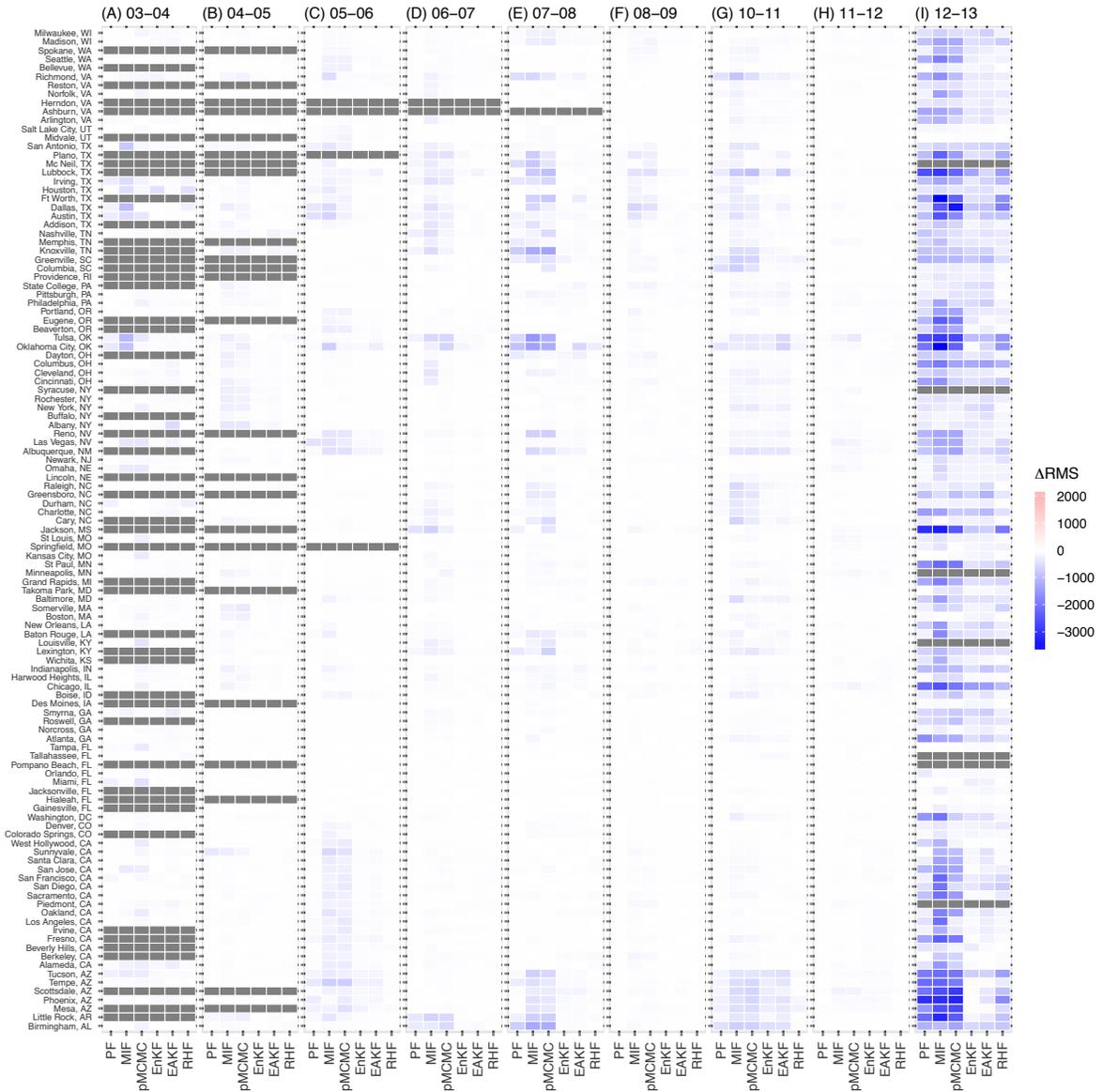

**Figure 4.** Comparison of model variable and parameter estimates made using the unmodified filters and the SR modified filters in which the state space of variable *S* was re-probed. (A) Estimates of the number of susceptible persons *S* at the epidemic onset (defined as the week with the maximum *S*). (B) Estimates of basic reproductive number $R_0$ at the epidemic onset. (C) Estimates of the effective reproductive number $R_e$ at the maximum epidemic forcing (defined as the week with the maximum $R_e$). (D) Estimates of the infectious period *D* at the maximum epidemic forcing. Estimates are shown segregated by season. The box and whisker plots represent the distribution of parameter values across all U.S. cities included in this study. The horizontal thick line is the median, the box edges are the 25th and 75th percentiles, the whiskers span the full range, excluding points (dots) deemed outliers.

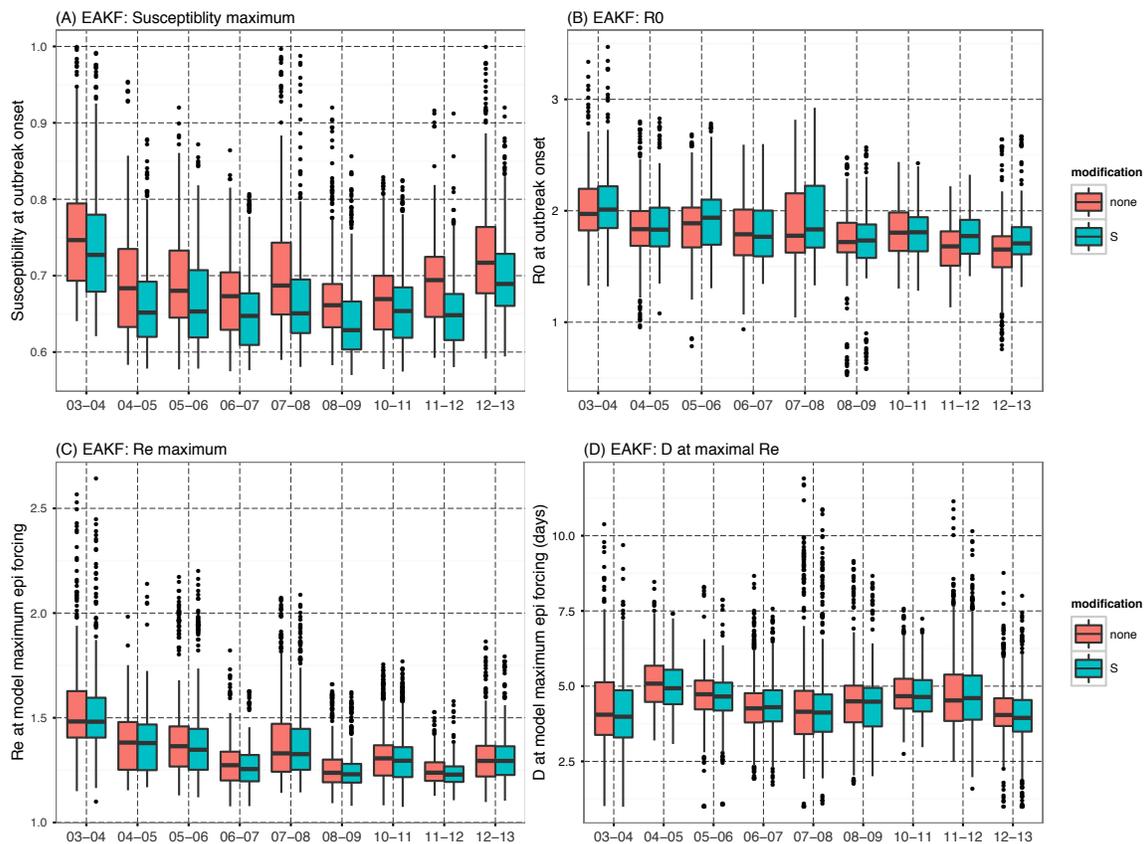

**Figure 5.** Comparison of outbreak peak timing prediction accuracy. The y-axis shows the accuracy, averaged over all cities for each season or over all 9 seasons (last row, shown as 'Epi03-13'), for each week of the year when the forecast was made (x-axis). Numbers greater than 52/53 are weeks in the next year, e.g., Week 54 is the first week in 2004 and 2008 in the 2003-04 and 2008-09 seasons, as Years 2003 and 2008 had 53 weeks, and is the second week in the rest of the seasons. The grey vertical line indicates the week most frequently observed as the peak for that seasonal outbreak among all cities.

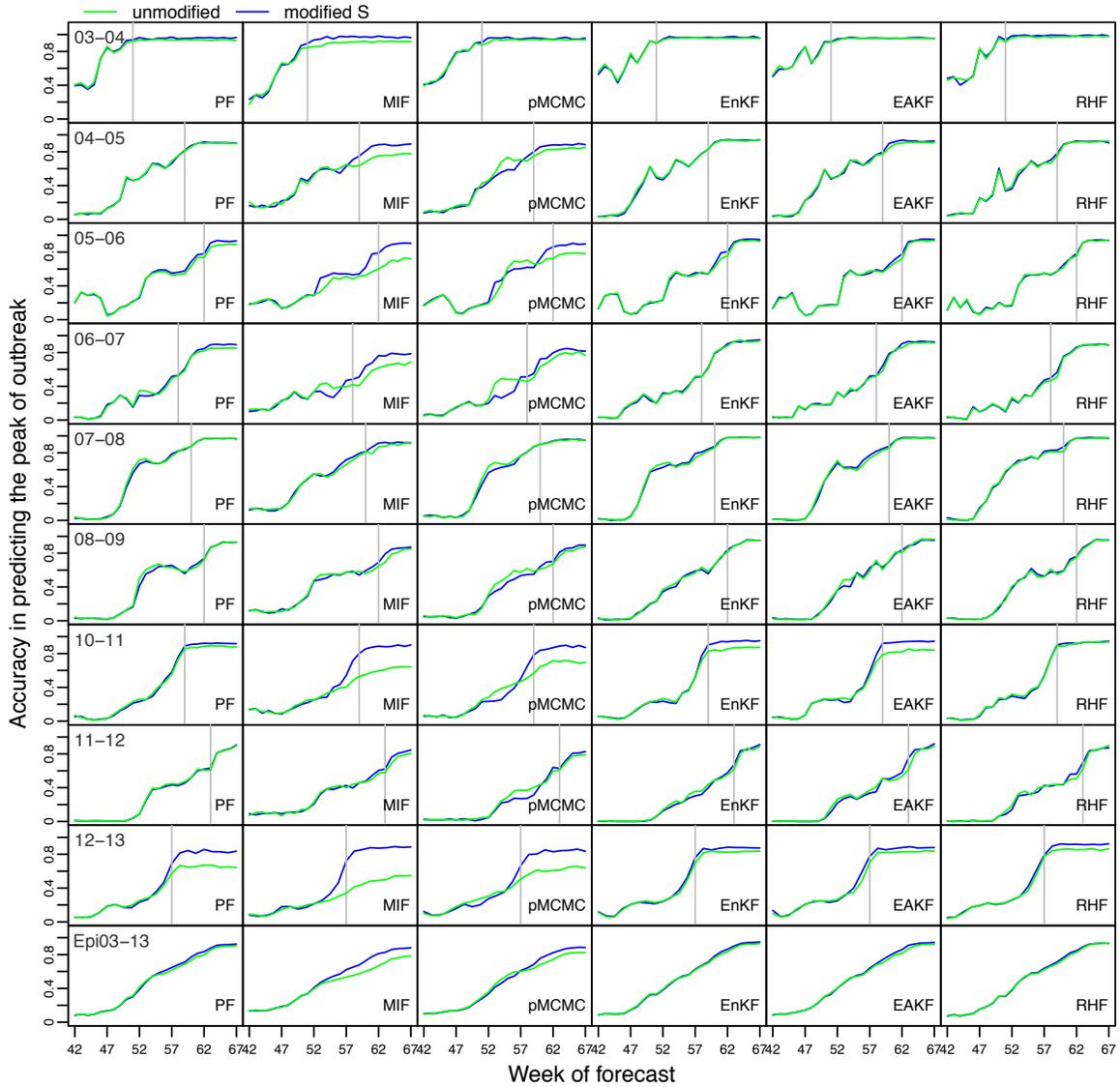

**Figure 6.** Difference in outbreak peak timing prediction accuracy (ΔAcc) summed over 9 epidemic seasons for each city, made at 4 weeks prior to through 4 weeks post the actual peak for the six filters (A-F). ΔAcc is calculated as the accuracy produced with the SR filter minus that by the corresponding unmodified filter. The x-axis shows the number of weeks relative to the observed peak; positive numbers are before the peak and negative ones are after the peak.

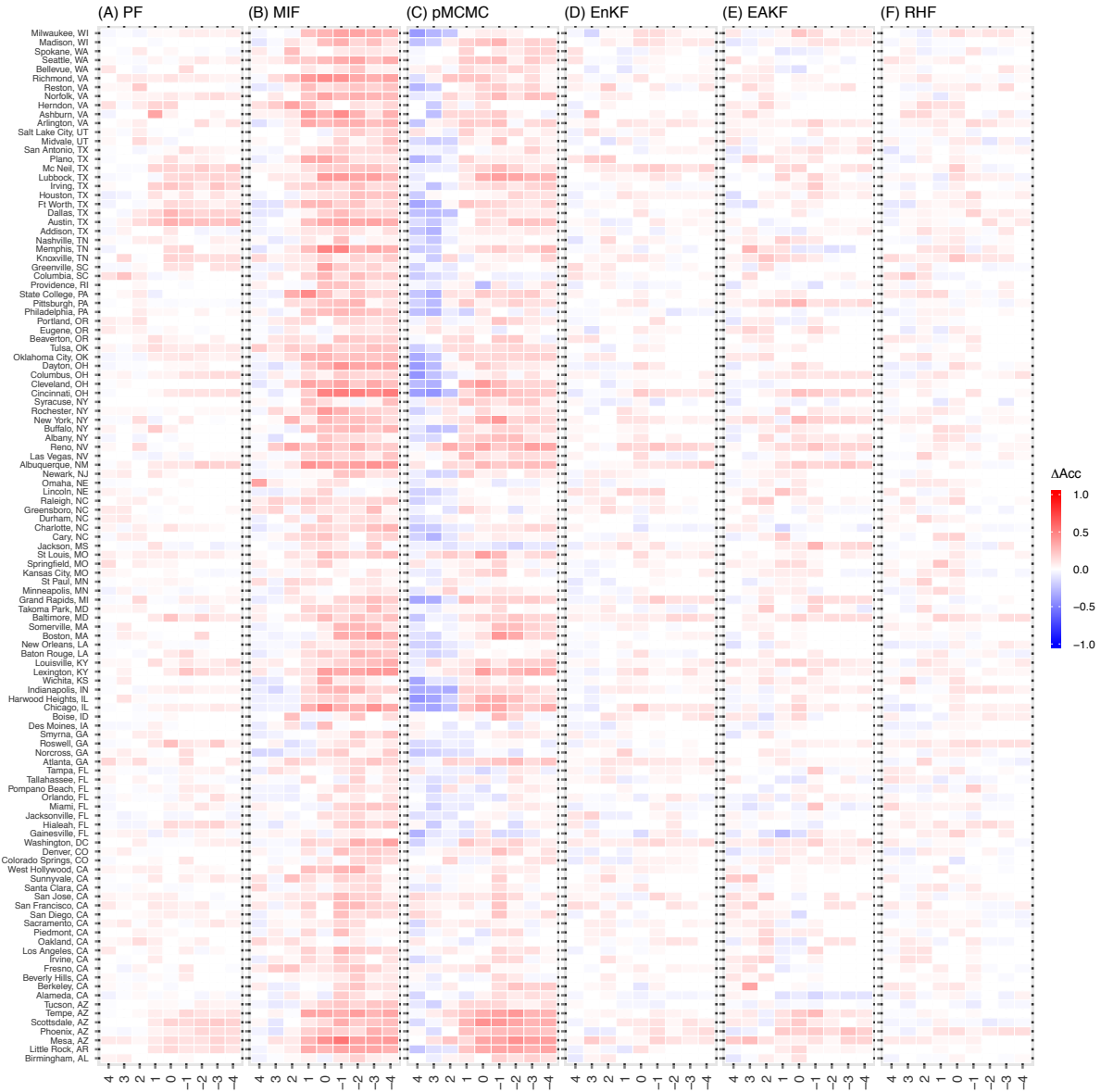

# Supplemental Material

## A simple modification for improving inference of non-linear dynamical systems

Wan Yang, Jeffrey Shaman

We tested the proposed modification, termed space reprobing (SR), through simulation and forecast of influenza seasonal outbreaks. The setting of the problem is the same as described in [1]. The SR modification is applied to six filters, i.e., three particle filters—a particle filter with conditional resampling and regularization (PF) [2], maximum likelihood estimation via iterated filtering (MIF) [3], and particle Markov chain Monte Carlo (pMCMC) [4]—and three ensemble filters—the ensemble Kalman filter (EnKF) [5], ensemble adjustment Kalman filter (EAKF) [6], and rank histogram filter (RHF) [7]. For more details on each of the six filtering frameworks, refer to [1]. Each filter is used in conjunction with the same Susceptible-Infected-Recovered-Susceptible (SIRS) epidemiological model and observations of influenza incidence termed ILI+. The SIRS model includes 2 model variables, the number of susceptible persons, $S$, and number of infected persons, $I$, and 4 model parameters, the immunity period, $L$, the infectious period, $D$, and the maximum and minimum reproductive numbers, $R_{0max}$ and $R_{0min}$.. For more details on the SIRS model, refer to Shaman et al. 2013 [8].

In the following supplemental text, we first present the SR algorithm and illustrate the mechanisms of the SR modification for both the particle and ensemble filters. We then report results from preliminary testing of the SR modification applied to the susceptibility variable. These analyses are used to determine settings for space re-probing, as well as the optimal particle size for SR particle filters. In addition, we test the SR on other model parameters for comparison with SR applied to the $S$ variable.

## 1. Modification proposed in this study

The SR modification replaces a particular model variable/parameter for a few particles/ensemble members, at the end of every prediction-update filtering cycle. This same SR modification is applied to all six filters tested in this study. The modified filtering process is illustrated in Table S1, for the PF. In the algorithm, $\boldsymbol{x}$ (in bold) is the state, $w$ is the weight (i.e. the cumulative likelihood for each particle), and $z$ is the observed variable; the superscript $i$=1,…N, is the index of particle number, and the subscript $k$=0,..,K is the time index, where $k$=0 is the initial time step. For instance, $\boldsymbol{x}_k^i$ is the state of $i$th particle at the $k$th time step, and $w_k^i$ is its weight. $\pi(a|b)$ denotes the probability distribution of $a$ conditioned on $b$. For the SR step, M particles ($j$=1,..., M) are randomly selected, and the model variable or parameter chosen to be reprobed, $x_{reprobed}$, is re-assigned a new randomly selected value from a uniform distribution of values for each of the M particles.

## 2. Mechanisms of the SR modification for the particle and ensemble filters.

In general, the SR modification expands the spread of the particles/ensemble such that the filtering framework, either based on a particle or ensemble filter, can more reliably contain trajectories capable of representing new system dynamics. However, it works somewhat differently for the particle and the ensemble filters. To illustrate how it works on these two families of filters, we compared the prior and posterior distributions of the susceptible level, *S*, and modeled ILI+, as generated by the unmodified or the SR modified filters.

Ideally, a particle filter should include particles that saturate the full span of state space. This is roughly true at the early stage of a filtering process; however, the number of unique particles decreases over time as particles deemed unlikely based on past observations are assigned reduced weight, and ultimately eliminated during resampling, while those deemed more likely are given greater weight and increasingly replicated. Consequently, PFs can suffer particle impoverishment, in which only a few unique particles exist and only a limited portion of state space is explored. Regularization, in which the replicate particles are randomly varied slightly from the original resampled particle, can be used to eliminate particle redundancy [2], but this process does not greatly increase the breadth of state space being spanned. Under these circumstances, the filter is no longer exploring the full span of state space and filter performance could degrade if a change in system dynamics leads to an alteration of optimal sub-state space (Fig. S1A and S1B, left panels).

The SR modification enables the particle filter to continually explore a wider span of state space and prevents particle impoverishment (Fig. S1A and S1B, right panels). This effect is achieved by re-assigning a small fraction (e.g., 2%) of particles to values over any possible region of the state space. If any of these probing particles turns out to better represent the system during or following a dynamical shift, the filter would assign those particles increased weight and magnify those trajectories; if not, the filter would assign them reduced weight and eliminate them at the next assimilation checkpoint. Further, since only a small fraction of particles are used for re-probing, the filter can maintain function by the unaltered, majority stream of particles.

For ensemble filters, the SR modification broadens the distribution for one or more of the variables or parameters. Even if these re-probed variables/parameters are unobserved, the broadened distribution alters their covariance with the observed variable(s). This change can ultimately feedback and alter the spread of the observed variable, which reduces confidence in the prior, places greater confidence in the observation, and prevents system divergence (Figure S2). Altered covariance of the observed variable(s) with the other unmodified variables/parameters may then provide further feedback.

## 3. Testing the SR modified filters

We first explored the SR modification on two particle filters, the MIF and pMCMC. The MIF and pMCMC assume that model parameters are constant over time, as opposed to time-varying, as treated by the other four filters. As a result, these two filters are the least flexible and can only alter model variables to reflect a potential alteration of system behavior (e.g., change in dominant circulating virus strain).

Space re-probing was performed on the unobserved susceptible variable ($S$).  A number of preliminary analyses were performed to determine:

(1) When to apply the modification, e.g., at every prediction-update filtering cycle or triggered only by specific events?
(2) How to apply the modification, e.g., through random modification of particles or selective modification?
(3) How many particles to replace?
(4) The range of values to use when modifying S?

Because changes to model system dynamics (e.g. appearance of a new dominant influenza strain) are not always predictable, a reactive space re-probing method in which the SR modification is triggered on the fly by certain defined events was found to be less desirable.   In particular, triggered re-probing would require continuous and comprehensive screening for all potential scenarios in which SR might be warranted.  Such comprehensive screening is likely not possible but would rather miss some events that warrant re-probing; furthermore, the screening process would necessitate additional IF-THEN operations within the SR algorithm, which would increase computational cost.  As a consequence, we simply applied the SR modification at every prediction-update cycle.  As discussed in the main text, this strategy enables the filter to quickly capture a dramatic change in the system (e.g., a second rise in influenza incidence).  On the other hand, if there is no shift, the modified particles would be weighted down at the next update check point and would not disturb the filtering process.

Two modification strategies were tested: 1) selective SR in which particles with lowest weight were replaced; and 2) random replacement.  Performance of the SR modified MIF with selective replacement was inconsistent. In contrast, the SR modified MIF with random replacement improved the performance more consistently over time (Fig S3A).  Both strategies worked for the pMCMC; however, simulations with random replacement often more effectively represented and forecast erratic outbreaks (due to multiple influenza strains) (Fig S3B). We thus used random replacement in subsequent tests.

In testing the fraction of particles to re-probe, we found that replacing 1% to 4% of the particles improved the performance of the MIF and the pMCMC (Fig S4).  Re-probing with 1% of particles had a less marked effect than re-probing with 2% or 4% of particles. However, greater than 2% particle re-probing did not show obvious increased benefit.  To ensure sufficient re-probing of particles while avoiding any disturbance to the filtering process, we therefore used a replacement level of 2% of particles in all further tests.

Finally, we explored the range of $S$ values to be used during re-probing.  In all instances, S values are randomly drawn from a uniform distribution with a defined range.  We used a fixed lower range bound of 50% susceptibility (i.e. 50% of the population size) to reflect a minimum increase of susceptibility should a new dominant strain appear.  Three different upper susceptibility bounds, 60%, 70%, or 80%, were tested. For both the MIF and pMCMC, replacement with all three upper bounds generates predictions close to the epidemic curve; however, use of the higher upper bounds, i.e., 70% or 80%, tended to yield somewhat unrealistic estimates of key model parameters (i.e. $D$ and $R_{0max}$). We thus used 60% susceptibility as the upper bound for the epidemic seasons.

## 4. Test the particle size

In our previous study [1], we found that the unmodified particle filters tested here, i.e., PF, MIF, and pMCMC, required 10,000 particles to achieve consistent inference, whereas an ensemble size of 300 was sufficient for performance of the three ensemble filters. Here we test whether the SR modified particle filters could perform well with fewer particles. As in Yang et al. [1], we ran SR-modified forms of the three particle filters with 300, 3000, or 10000 particles using historical ILI+ time series for Atlanta, Boston, Chicago, Los Angeles, New York, and Seattle. Increasing the particle number from 300 to 3000 significantly improved performance (lower RMS error or higher correlation, pairwise paired t-test, 1 sided, p<0.05). Increasing the particle number to 10000 did not significantly improve performance further (Figure S5). We thus used 3000 particles for all the SR particle filters.

## 5. Test the SR modification on other parameters.

Of the four parameters in our SIRS model, $D$ and $R_{0max}$ are more important determinants of within-season outbreak characteristics (Shaman and Karspeck, 2012 [9]). We thus tested whether SR modification of these two parameters, instead of the $S$ variable, would similarly improve filter performance.

Tests were performed in which 2% of particles/ensemble members were randomly replaced for either $D$ or $R_{0max}$. Values were drawn from uniform distributions: unif (2, 7) for re-probing of $D$, and unif (2, 5) for $R_{0max}$. We tested these versions of the SR modification using the PF and EAKF by simulating historical ILI+ time series and forecasting outbreak peak timing, as done for the SR modification with $S$.

(1) Fitting epidemic curves
We ran the PF and EAKF 5 times to fit historical ILI+ time series for each of 9 epidemic seasons during 2003-04 through 2012-13 (excluding the 2009 pandemic) for 115 cities in the U.S. Both filters were run either with the SR modification to $S$, $D$, or $R_{0max}$, or without any modification. Each seasonal simulation began from Week 40, of a given year and ended at Week 39 of the next year. For the 2012-13 season, only 25 weekly ILI+ records collected from Week 40 of 2012 were available at the time of this study; the simulation was done only for those 25 weeks (i.e., from Week 40 in 2012 to Week 12 in 2013). As influenza activity was early in that season, those 25-week ILI+ records nevertheless cover the peak of all cities for that season. The RMS error for each run was calculated, and then averaged over the 5 repeated runs. ΔRMS was calculated as the mean RMS from an SR modified filter minus that from the corresponding unmodified version. The cumulative density distributions for ΔRMS are shown in Fig. S6. All versions of the SR modification significantly improved the PF and EAKF (lower RMS error, p<0.05).

(2) Retrospective forecast

As in Yang et al. [1], we ran the PF and EAKF to forecast the peak timing of influenza outbreaks. A forecast includes a training period which assimilates past ILI+ records up to the time of prediction, and a prediction step which takes the optimized model variables and parameters inferred by the training as initial conditions and integrates the SIRS model forward to the end of each season. This training and forecast procedure was performed on a weekly basis for 25 weeks beginning with Week 42 for all seasons and cities. Each filter was run either with or without the SR modification (to $S$, $D$, $R_{0max}$, or a combination of any two or three of them). The simulated outbreak peak was calculated as the week with the highest modeled incidence over the entire season (i.e., training plus prediction) based on the mean trajectory of all particles/ensemble members. A simulated peak within ±1 week of the observed ILI+ peak was deemed accurate. The RMS error was calculated between the predicted ILI+ time series over the entire season and the observed ILI+ time series. The entire forecasting procedure was repeated 5 times.

Modifying $D$ or $R_{0max}$, especially $R_{0max}$, had an effect similar to modifying $S$ (Figs S7 and S8). While a single variable/parameter is explicitly modified, it results in systematic adjustment to the other variables/parameters (Figs S7 and S8). For the particle filters, this adjustment is the result of the altered weighting for modified particles at the next assimilation checkpoint; for the ensemble filters, the systematic adjustment of the entire state is due to adjustment of the observed-unobserved variable/parameter covariance matrix. Modifications to $D$ or $R_{0max}$, or a combination of $S$, $D$, $R_{0max}$ produced similar improvements for both the PF and the EAKF (Figs S9 and S10). When compared with SR modification of $S$ or $R_{0max}$ only, modification of more than one variable/parameter did not significantly further improve filter performance.

**Table S1.** Pseudo code for an unmodified particle filter and the SR modified version

| Unmodified version (adapted from Algorithm 3 in Arulampalam et al. 2002) | SR Modified version |
|---|---|
| 1. Initialization: at k=0, $\{x_0^i, w_0^i\}_{i=1}^N$<br>$\quad where, w_0^i = 1/N$<br>2. FOR k=1:K (time steps)<br>   - FOR i=1:N (particles)<br>     • Draw $x_k^i \sim \pi(x_k \vert x_{k-1}^i, z_k)$<br>     [Note: this is to integrate the SIRS model forward 1 step, which generates a <u>prediction</u>]<br>     • Assign the particle a weight, $w_k^i$, according to:<br><br>$w_k^i \propto w_{k-1}^i \dfrac{\pi(z_k \vert x_k^i)\pi(x_k^i \vert x_{k-1}^i)}{\pi(x_k^i \vert x_{k-1}^i, z_k)} \propto w_{k-1}^i \pi(z_k \vert x_k^i)$<br><br>     assuming:<br>$\qquad \pi(x_k^i \vert x_{k-1}^i) = \pi(x_k \vert x_{k-1}^i, z_k)$<br>     [Note: this generates the <u>updated</u> weight/particles]<br>   - END FOR<br>   - Calculate total weight:<br>$\qquad t = sum[\{w_k^i\}_{i=1}^N]$<br><br>   - FOR i=1:N<br>     • Normalize: $w_k^i = t^{-1} w_k^i$<br>   - END FOR<br><br><br><br><br><br><br><br><br>   - Calculate $\widehat{N_{eff}}$ using:<br>$\qquad \widehat{N_{eff}} = \dfrac{1}{\sum_{i=1}^N (w_k^i)^2}$<br><br>   - IF $\widehat{N_{eff}} < N_T$<br><br>[Note: in this study, $N_T$=N/2]<br>     • Resample particles according to their weight<br>     • Set new weight to 1/N<br>   - END IF<br> END FOR (time steps) | 1. Initialization: at k=0, $\{x_0^i, w_0^i\}_{i=1}^N$<br>$\quad where, w_0^i = 1/N$<br>2. FOR k=1:K (time steps)<br>   - FOR i=1:N (particles)<br>     • Draw $x_k^i \sim \pi(x_k \vert x_{k-1}^i, z_k)$<br>     [Note: this is to integrate the SIRS model forward 1 step, which generates a <u>prediction</u>]<br>     • Assign the particle a weight, $w_k^i$, according to:<br><br>$w_k^i \propto w_{k-1}^i \dfrac{\pi(z_k \vert x_k^i)\pi(x_k^i \vert x_{k-1}^i)}{\pi(x_k^i \vert x_{k-1}^i, z_k)} \propto w_{k-1}^i \pi(z_k \vert x_k^i)$<br><br>     assuming:<br>$\qquad \pi(x_k^i \vert x_{k-1}^i) = \pi(x_k \vert x_{k-1}^i, z_k)$<br>     [Note: this generates the <u>updated</u> weight/particles]<br>   - END FOR<br>   - Calculate total weight:<br>$\qquad t = sum[\{w_k^i\}_{i=1}^N]$<br><br>   - FOR i=1:N<br>     • Normalize: $w_k^i = t^{-1} w_k^i$<br>   - END FOR<br>   <span style="color:blue">- **Space Reprobing (SR) step:**<br>     FOR j=1:M<br>     • Randomly select a particle for reprobing<br>     • Randomly select a new value, $x_{reprobed}$, for that particle from a defined uniform distribution of values.<br>     END FOR</span><br><br>   - Calculate $\widehat{N_{eff}}$ using:<br>$\qquad \widehat{N_{eff}} = \dfrac{1}{\sum_{i=1}^N (w_k^i)^2}$<br><br>   - IF $\widehat{N_{eff}} < N_T$<br><br>[Note: in this study, $N_T$=N/2]<br>     • Resample particles according to their weight<br>     • Set new weight to 1/N<br>   - END IF<br> END FOR (time steps) |

**Supporting Figures**

**Figure S1A** Comparison of the prior and posterior distributions of susceptible number, *S*, generated by the unmodified vs. SR modified PF. For the purpose of demonstration, we here performed resampling at every prediction-update cycle (as opposed to conditional resampling used in the actual runs for the simulation and forecast). In doing so, the weight of each particle is equal (i.e., $1/N$, where $N$ is the total number of particle) such that the distribution of the particles represents the distribution of the state. Both versions of the filter were run with 300 particles. Each line represents the trajectory of a particle. The prior distributions were generated by integrating the SIRS model up to the next assimilation checkpoint, and the posterior distributions were calculated post assimilation of the latest ILI+ record. Regularization was applied to both versions of the filter.

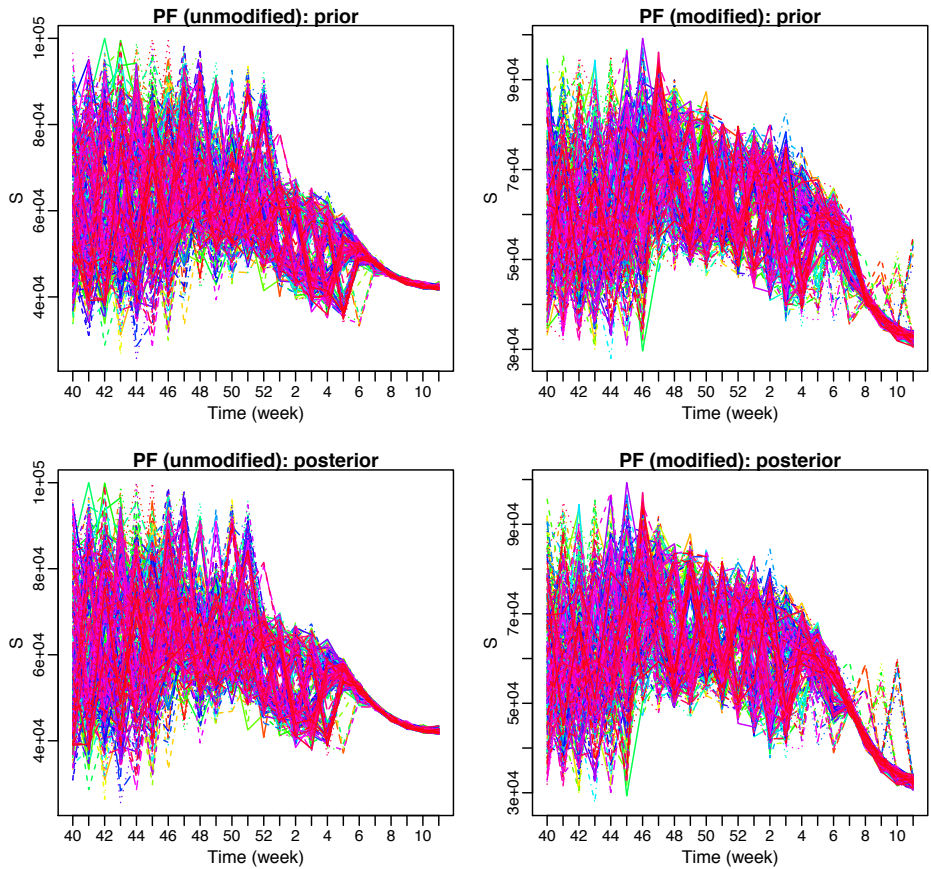

**Figure S1B** Comparison of the prior and posterior distributions of the observed variable, *ILI+*, generated by the unmodified vs. modified PF.

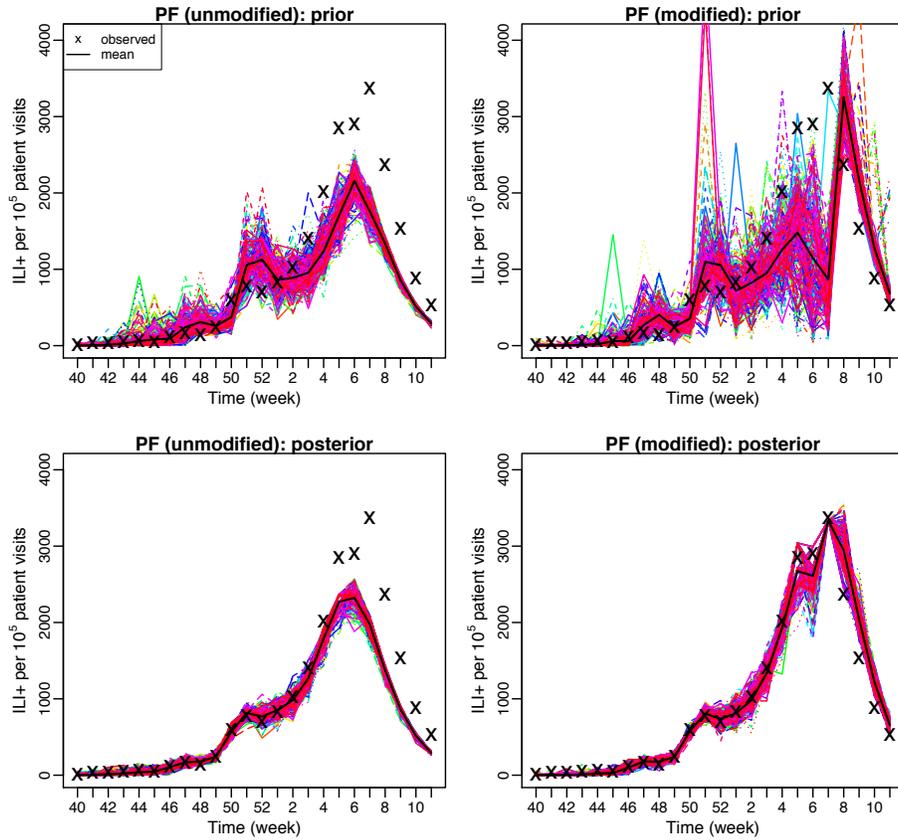

**Figure S2A** Comparison of the prior and posterior distributions of susceptible level, *S*, generated by the unmodified vs. SR-modified EAKF. Inflation was applied to both versions of the filter. Both versions of the filter were run with 300 ensemble members. Each line represents the trajectory of an ensemble member.

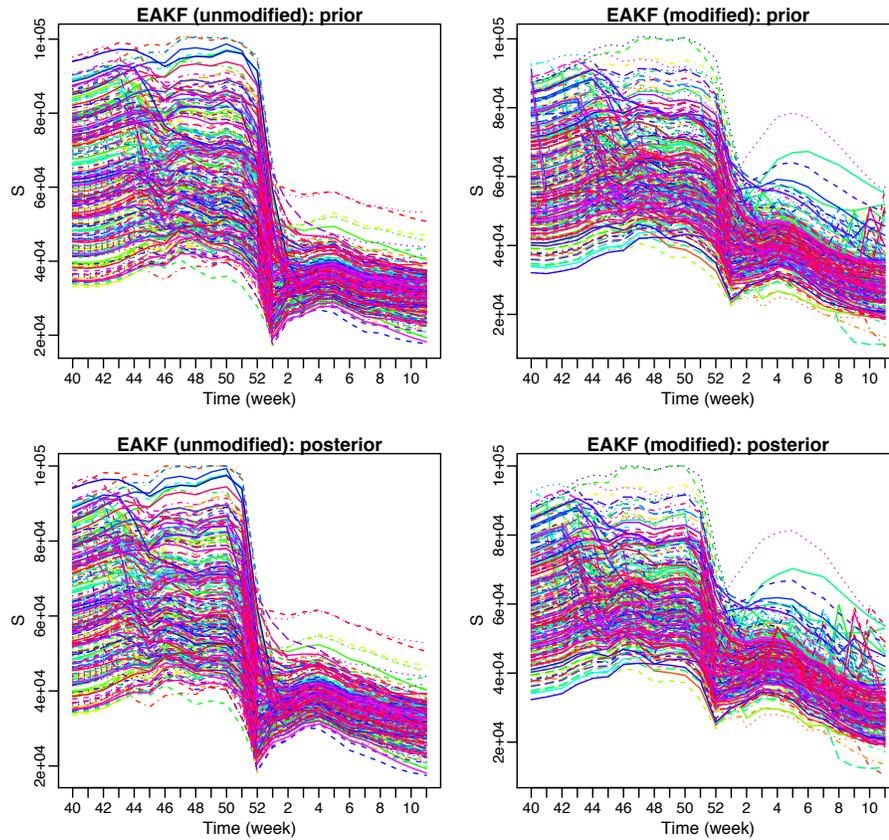

**Figure S2B** Comparison of the prior and posterior distributions of the observed variable, *ILI+*, generated by the unmodified vs. SR-modified EAKF.

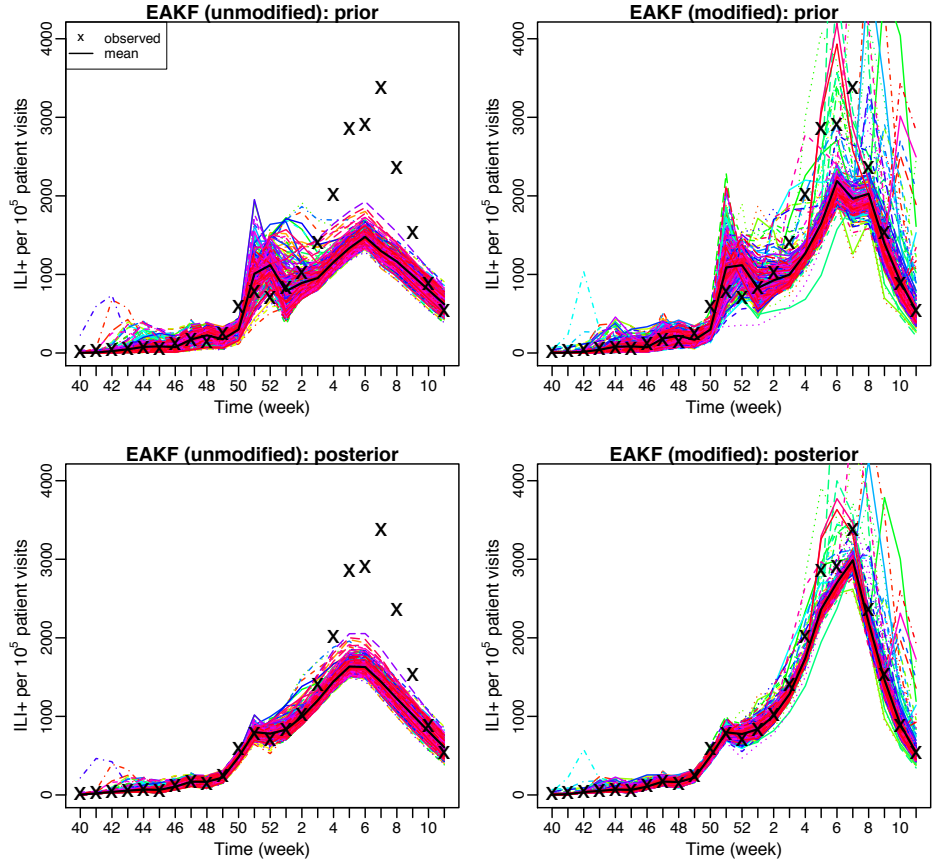

**Figure S3** Comparison of replacement methods for the MIF (A) and pMCMC (B). Each filter was run with the unmodified version, with 2% of the $S$ variable randomly replaced at each prediction-update cycle, or with 2% of $S$ with the lowest weighted particles replaced at each prediction-update cycle. 3000 particles were used in all versions of the filters. Solid lines are model fits over the training period and dashed lines are predictions by the filters. Phoenix, AZ had a bimodal outbreak in the 2010-11 season; each epidemic wave was probably caused by a different dominant strain, i.e. for the 2010-11 season in census division region 9, which includes Arizona, subtype A/H3 circulated early in season followed later by the 2009 H1N1 strain. Observations ('x') in red are available at the time of forecast and those in grey are unknown at the time.

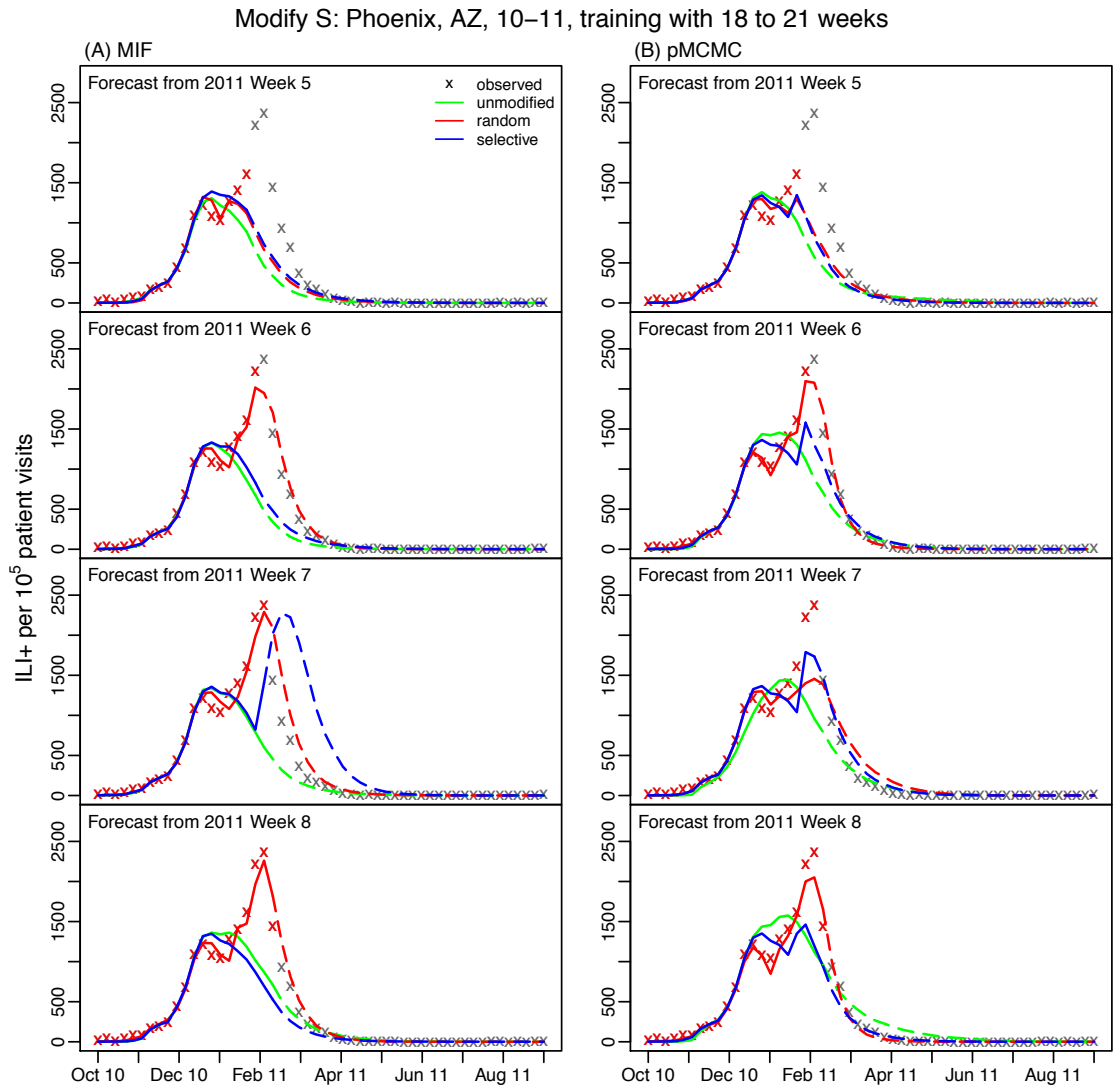

**Figure S4** Comparison of space re-probing for different fractions of particles using the MIF (A) and pMCMC (B). Each filter was run with 1%, 2%, or 4% of the *S* variable values randomly replaced at each prediction-update cycle. 3000 particles were used in all versions of these filters. Solid lines are model fits over the training period and dashed lines are predictions by the filters. Observations ('x') in red are available at the time of forecast and those in grey are unknown at the time.

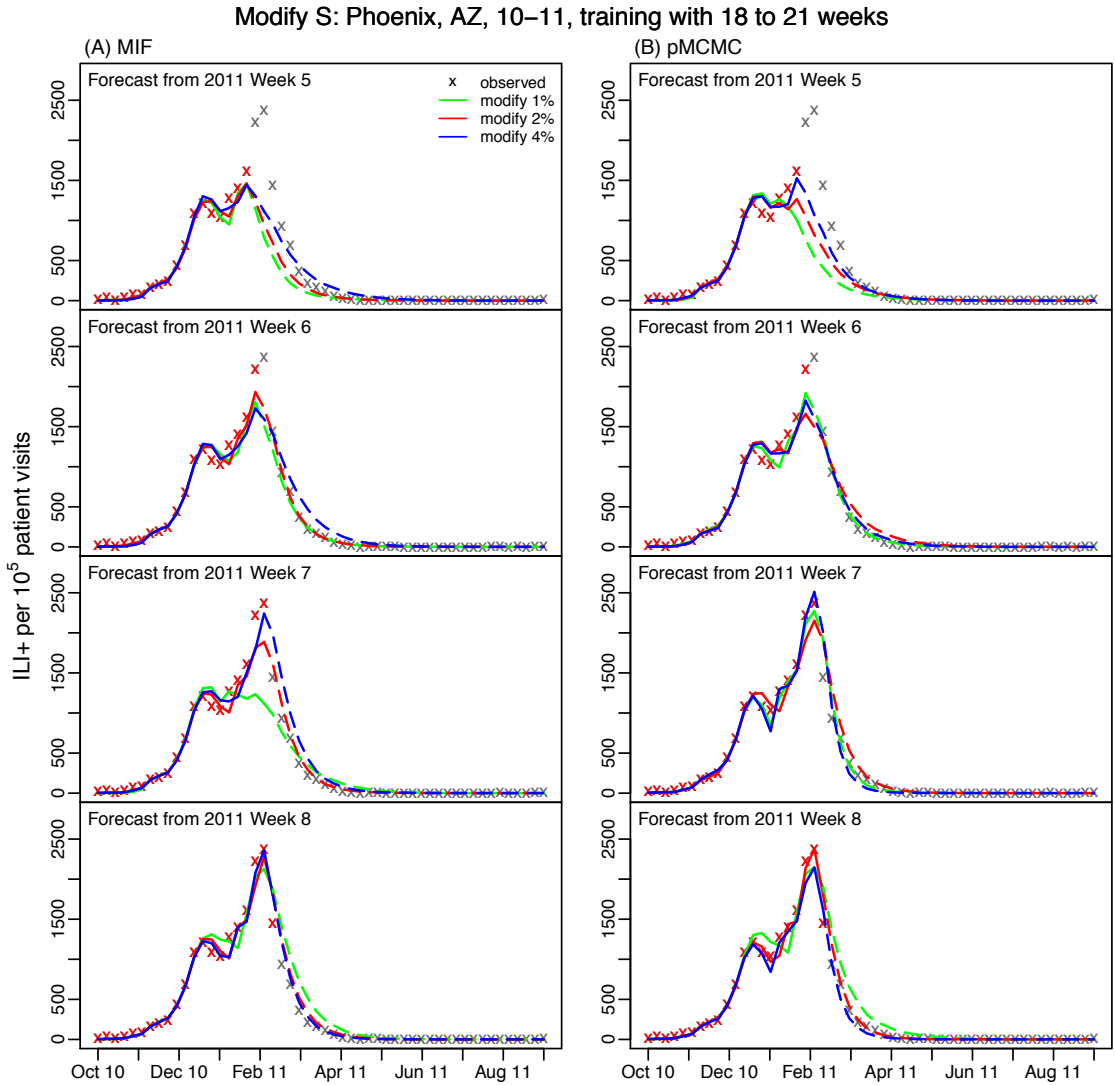

**Figure S5** Optimal particle size for the SR modified particle filters. The SR particle filters were run with either 300, 3000, or 10,000 particles, to simulate the historical ILI+ time series from 2003-04 to 2011-12 (excluding the pandemic seasons), for Atlantic, Boston, Chicago, Los Angeles, New York City, and Seattle. Each ILI+ time series simulation using each filter was repeated 5 times. The Root Mean Squared (RMS) error was calculated for each run. Boxplots of the RMS errors over the 5 runs show the performance of each filter with different particle sizes.

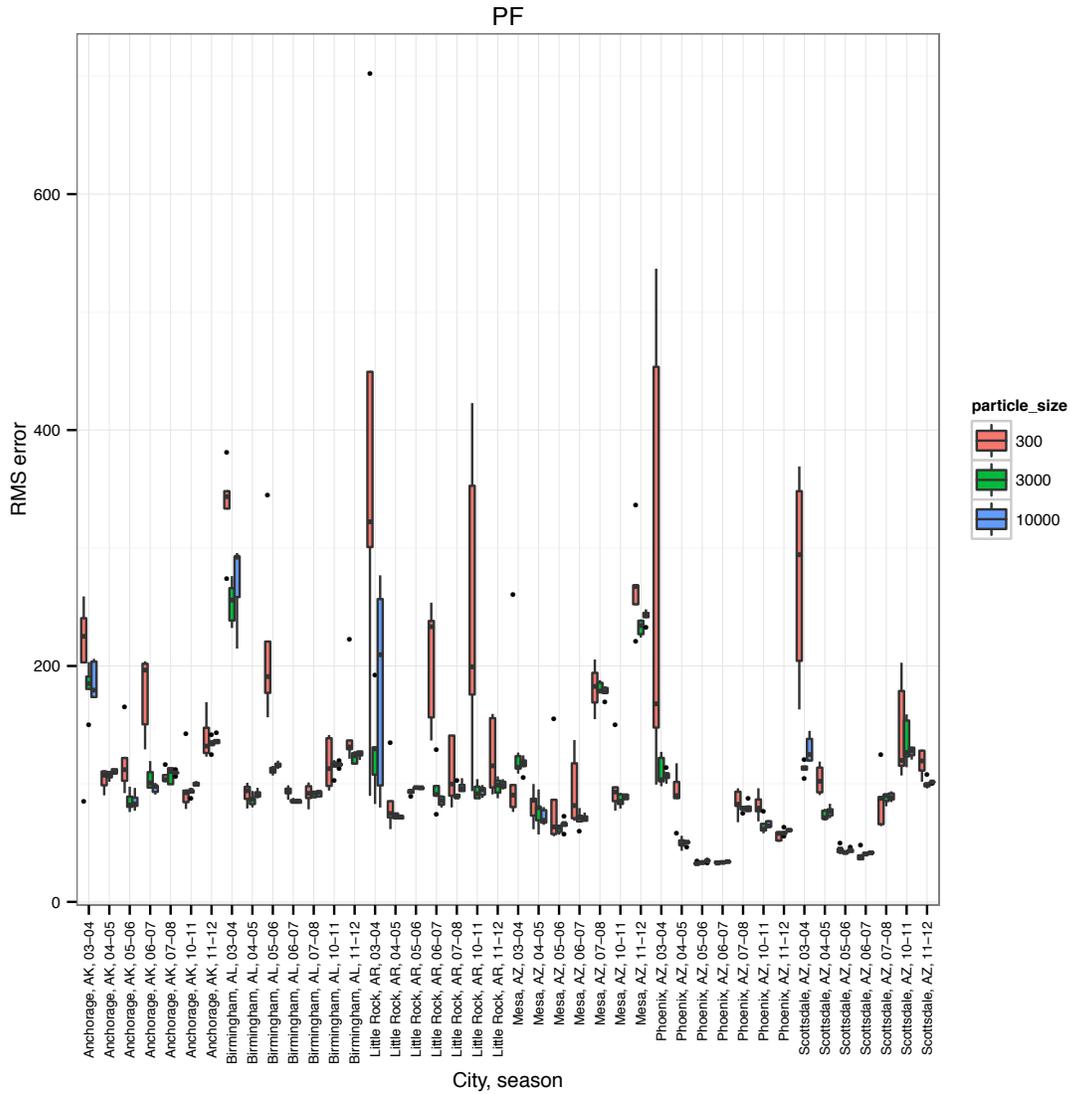

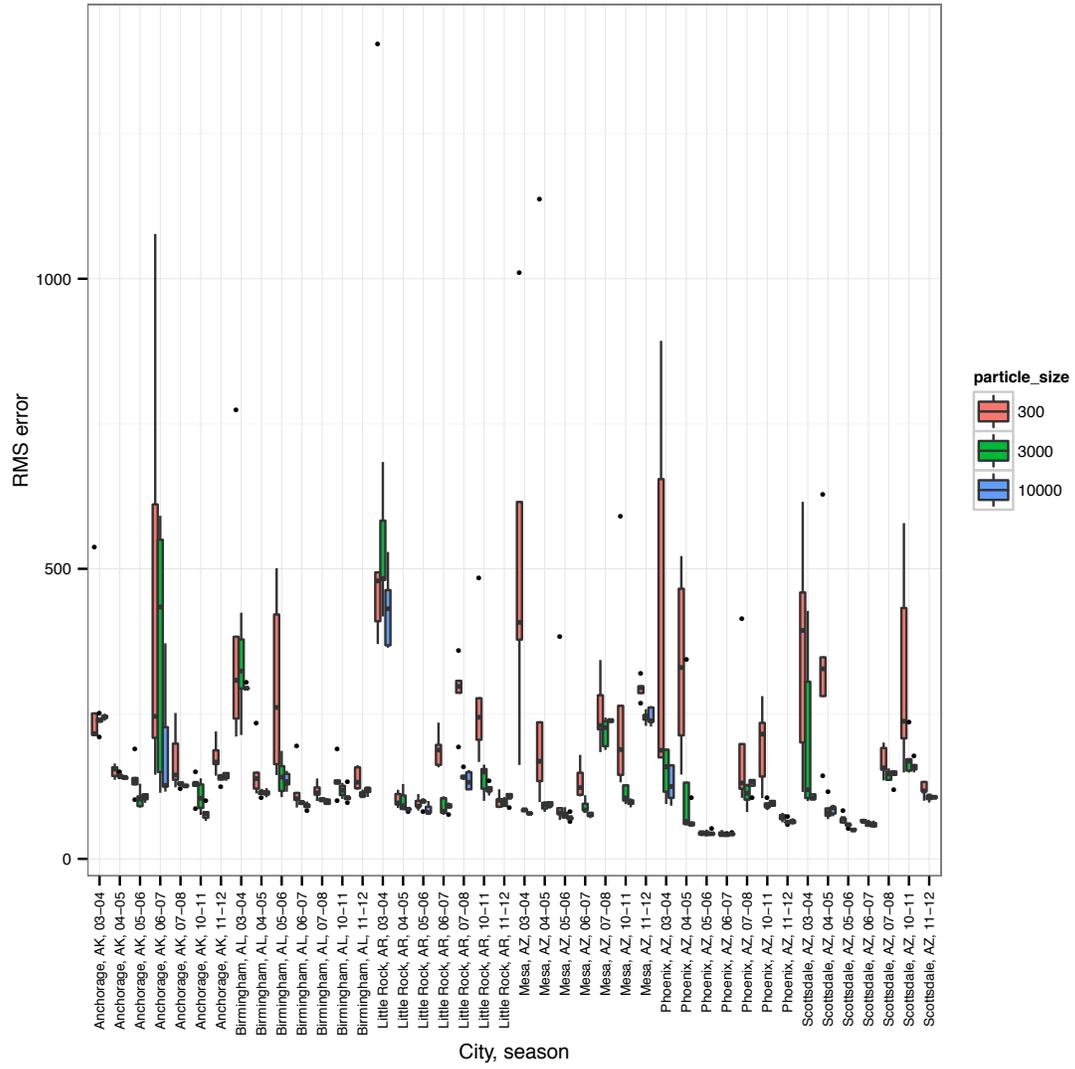

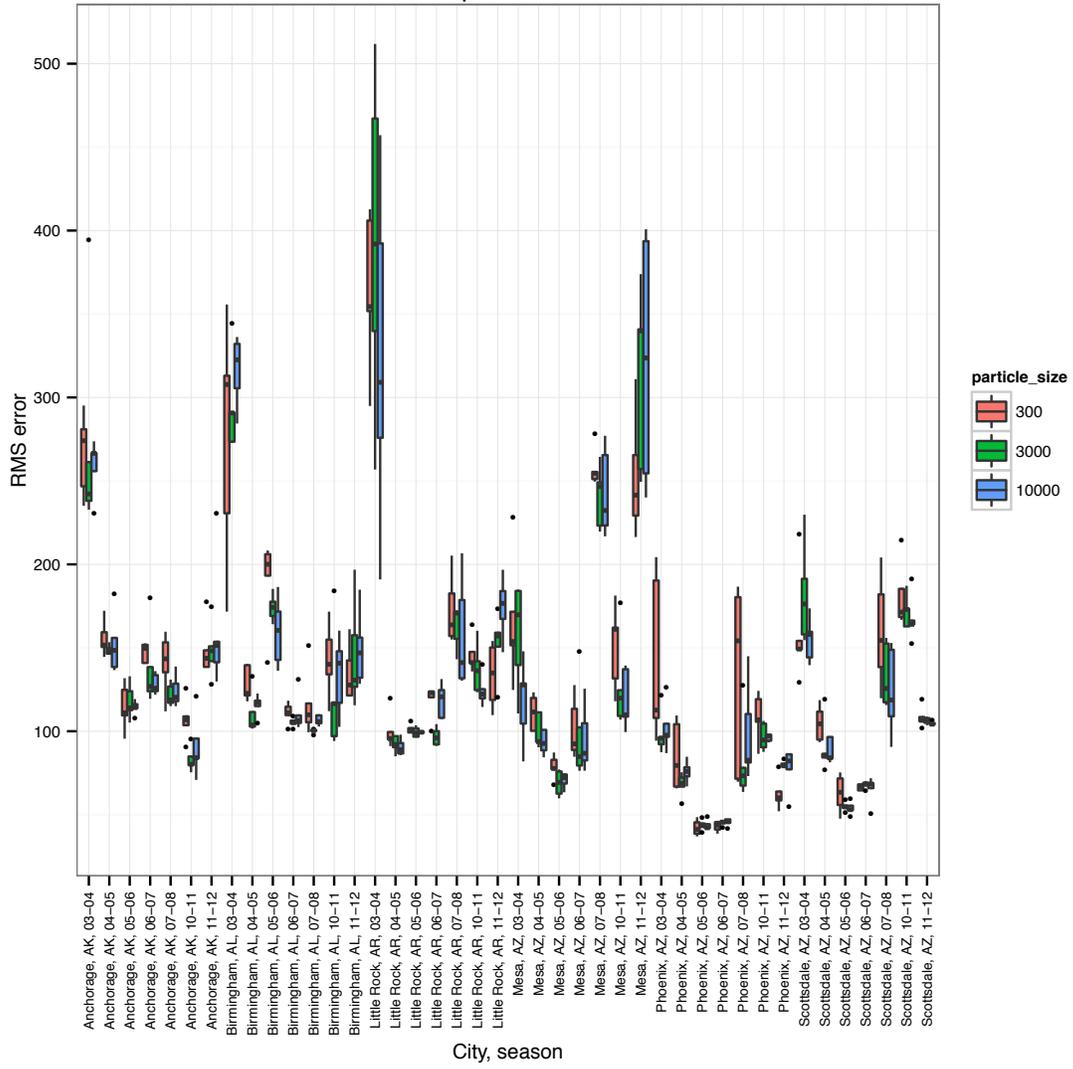

**Figure S6** Cumulative density distributions for the difference in Root Mean Squared (RMS) error between a filter with vs. without SR modification. The PF or EAKF was run repeatedly 5 times to fit historical ILI+ time series of 9 epidemic seasons during 2003-04 through 2012-13 for 115 U.S. cities, either with SR modification (to $S$, $D$, or $R_{0max}$) or without modification. The RMS error for each run was calculated, and then averaged over the 5 repeated runs. $\Delta$RMS was calculated as the mean RMS from a SR modified filter minus that from the corresponding unmodified filter.

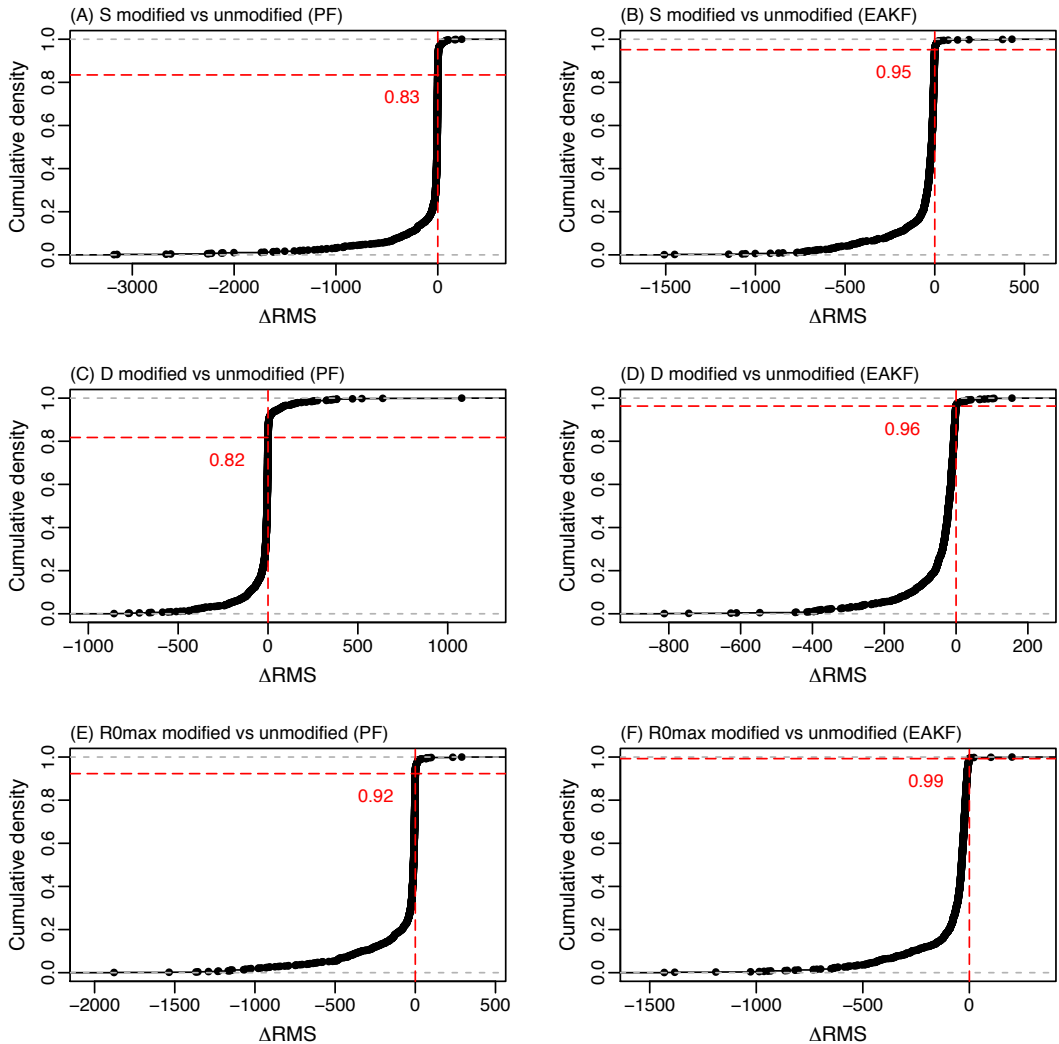

**Figure S7** Comparison of PF without SR modification and with SR modification of $S$, $D$, or $R_{0max}$. The PF was run in these four forms, to forecast the epidemic curve of 2010-11 outbreak in Phoenix, AZ with a training period of 18 to 21 weeks (the actual peak was at 2011 Week 7). Solid lines are simulated fittings during the training period and dashed lines are predictions. Although only one variable/parameter was SR modified, systematic changes to other variable/parameters ensued, as indicated by their diverse distributions. Observations ('x') in red are available at the time of forecast and those in grey are unknown at the time.

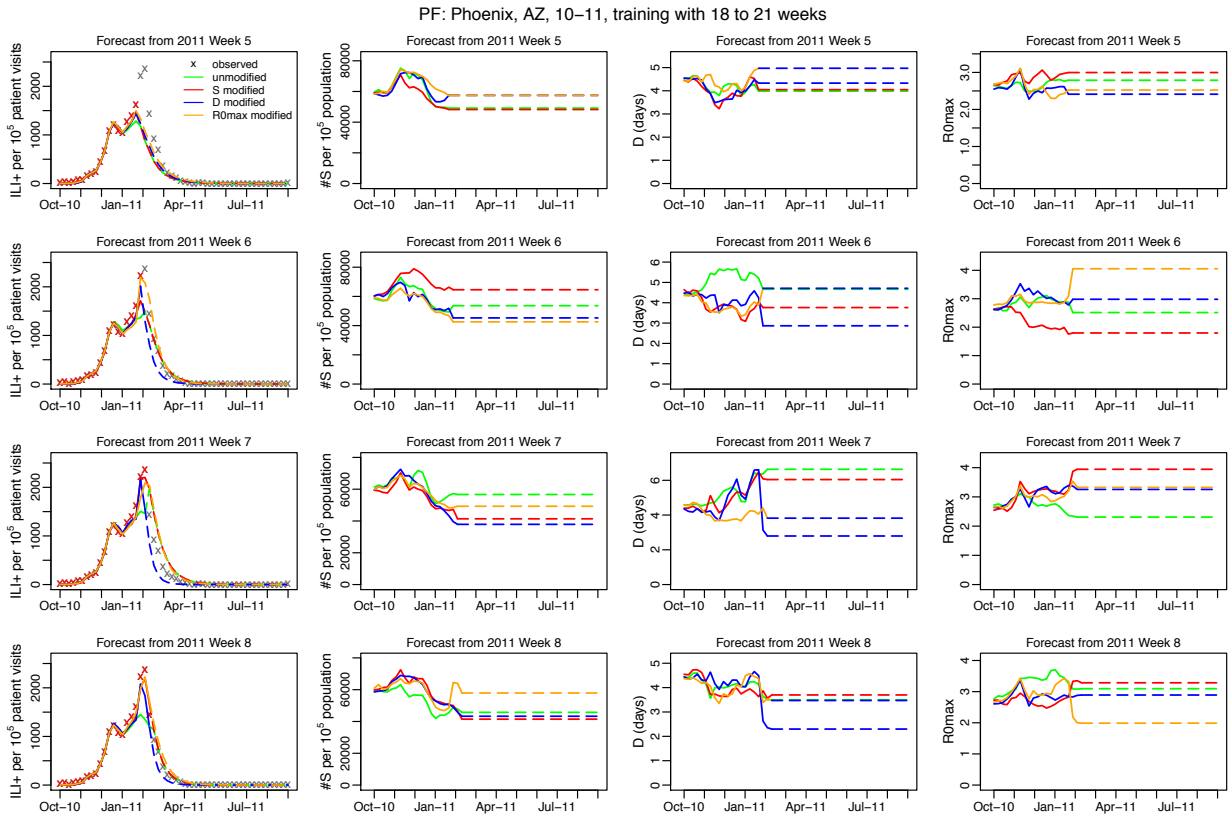

**Figure S8** Comparison of EAKF without SR modification and with SR modification of *S*, *D*, or $R_{0max}$. The EAKF was run in these four forms, to forecast the epidemic curve of 2010-11 outbreak in Phoenix, AZ with a training period of 18 to 21 weeks (the actual peak was at 2011 Week 7). Solid lines are simulated fittings during the training period and dashed lines are predictions. Although only one variable/parameter was SR modified, systematic changes to other variable/parameters ensued, as indicated by their diverse distributions. Observations ('x') in red are available at the time of forecast and those in grey are unknown at the time.

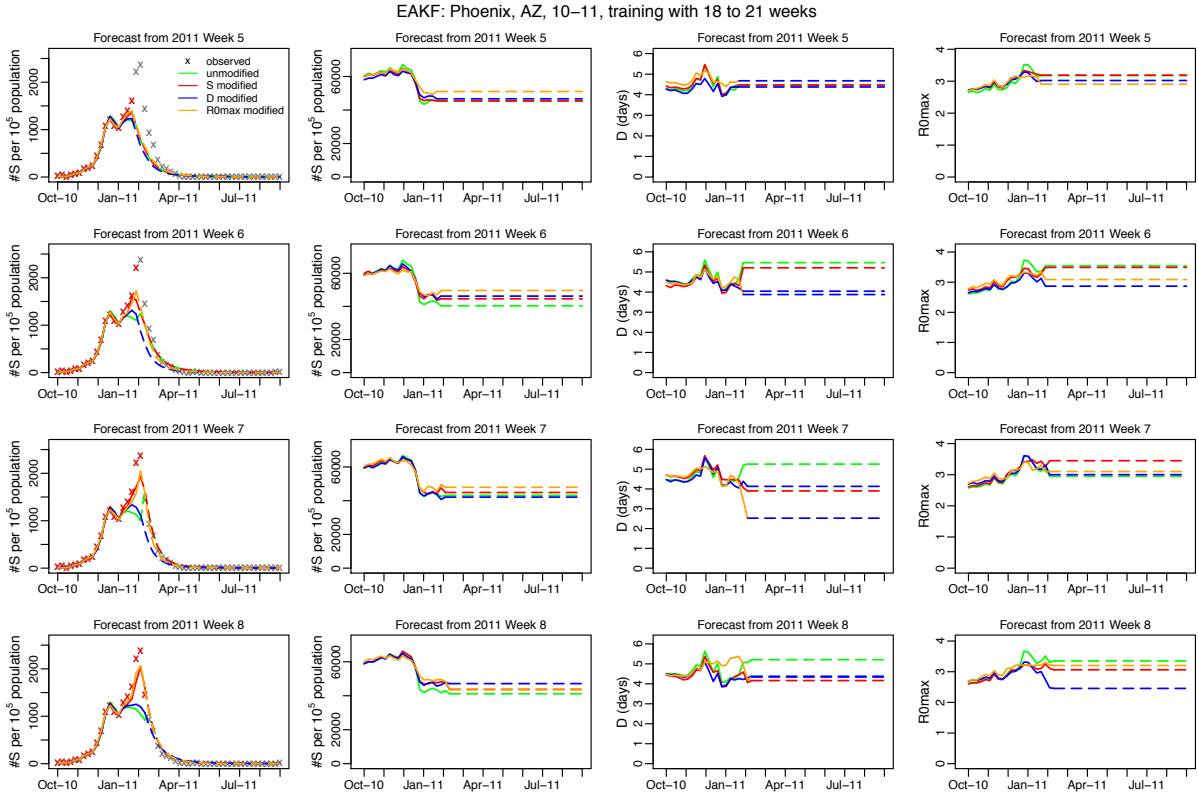

**Figure S9** Comparison of predicted outbreak peak timing accuracy for the PF. The PF was run without modification or with SR modification of *S*, *D*, or *R0max*, or any combination of the three, to retrospectively forecast the outbreak peak for each of the 9 epidemic seasons. Weekly forecasts generated during these 9 seasons were repeated 5 times for all U.S. cities. The overall accuracy (y-axis) for each week that forecast was made (x-axis) was then computed over all cities for each season (A-I) and over all 9 seasons (J).

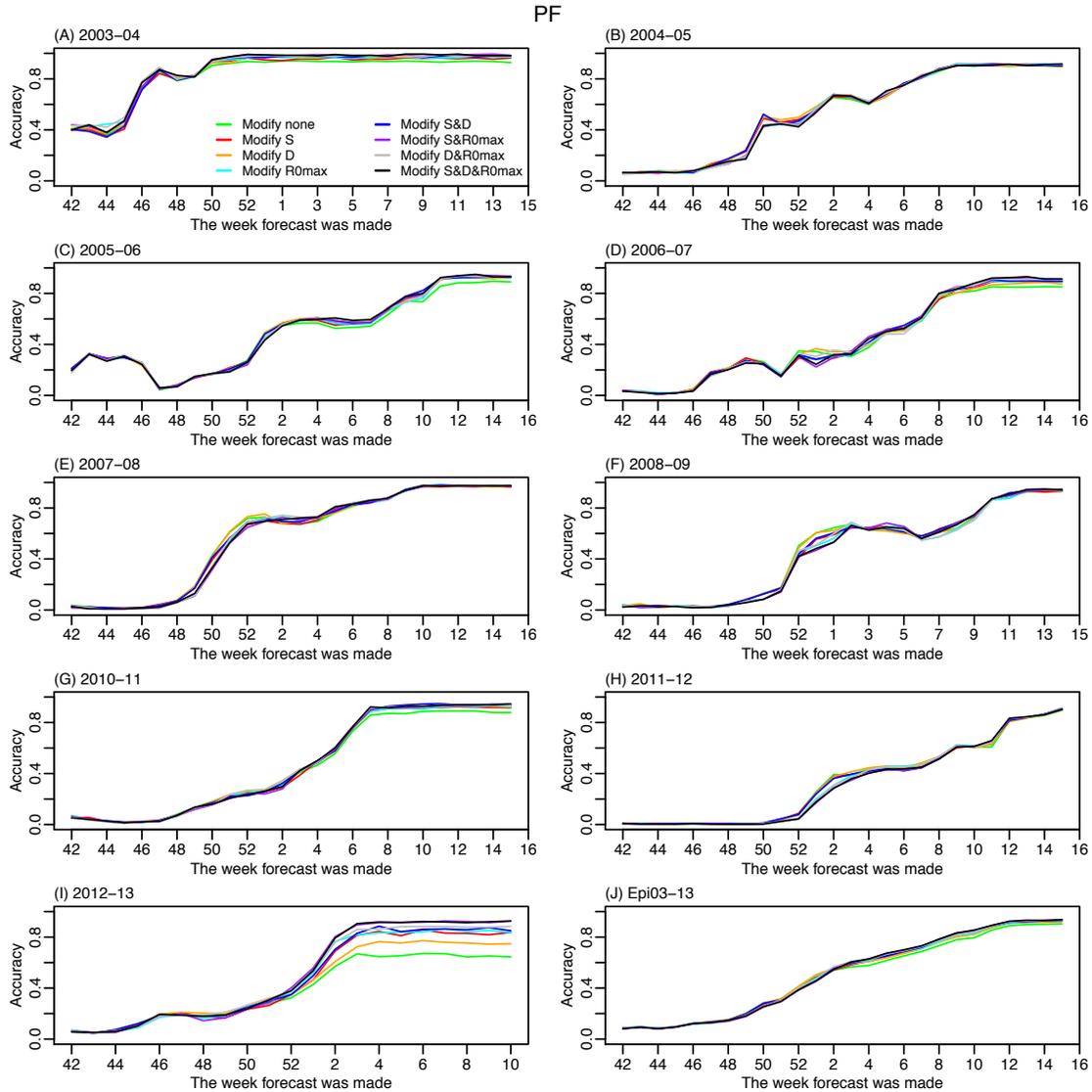

**Figure S10** As in Figure S9, but for the EAKF.

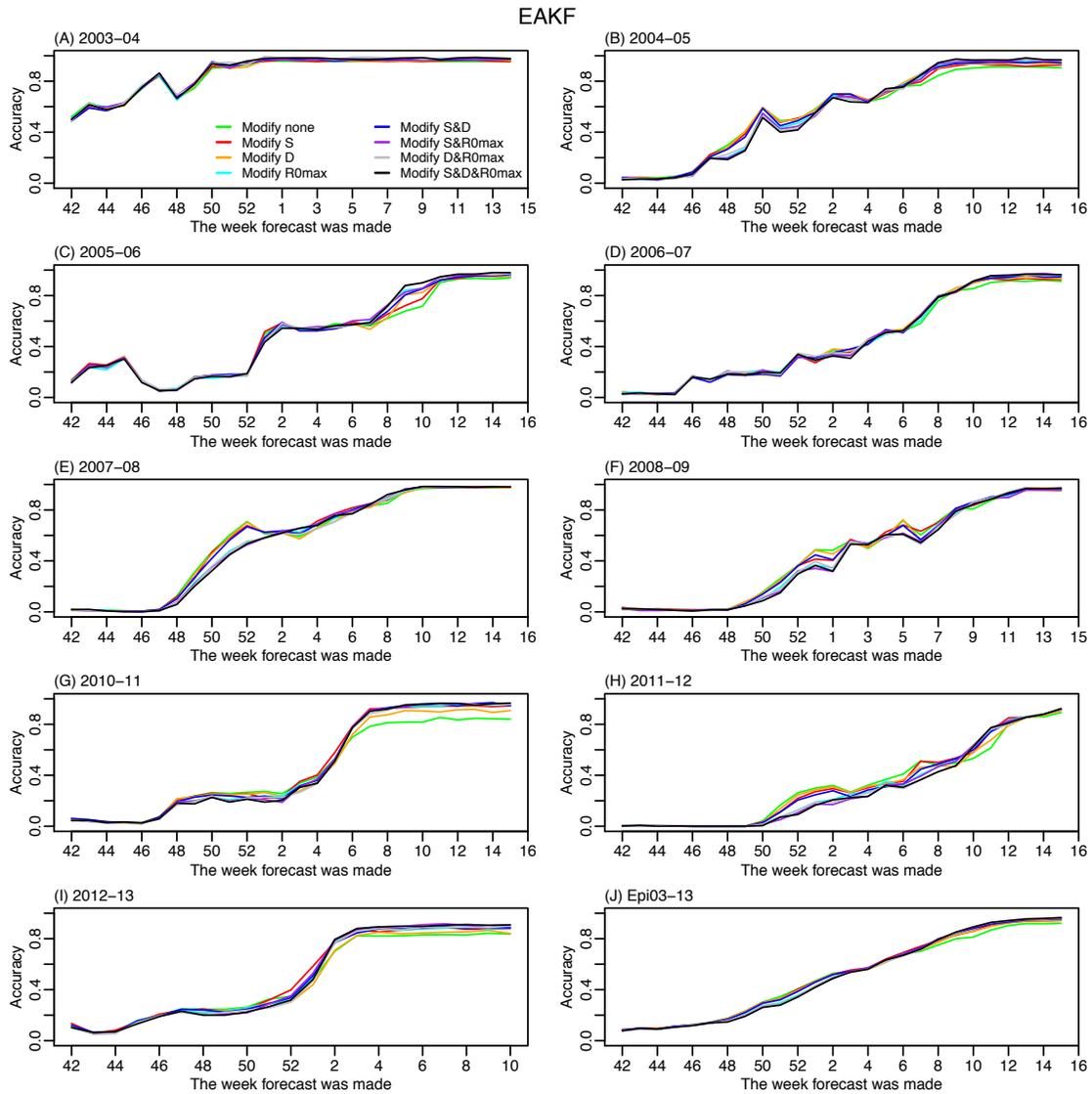

**Figures S11-S16** Difference in RMS error for the forecasts made with unmodified filters or filters with SR-modification of the $S$ variable. The RMS error for each forecast was calculated as the root mean squared difference in predicted and observed ILI+ records over each season. Results were then averaged over all cities for each week of forecast for each season (B-J) or over all 9 seasons (A).

PF

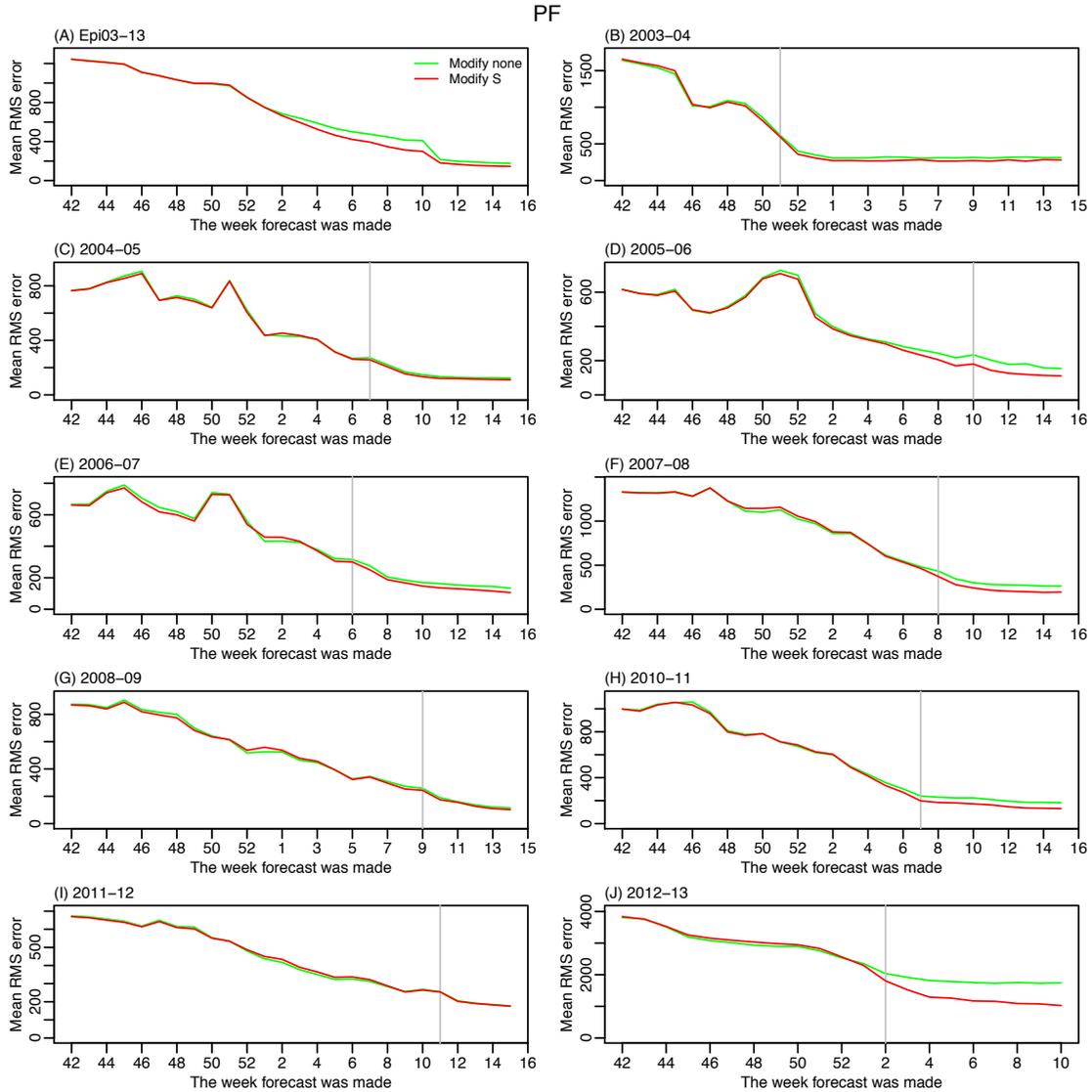

## MIF

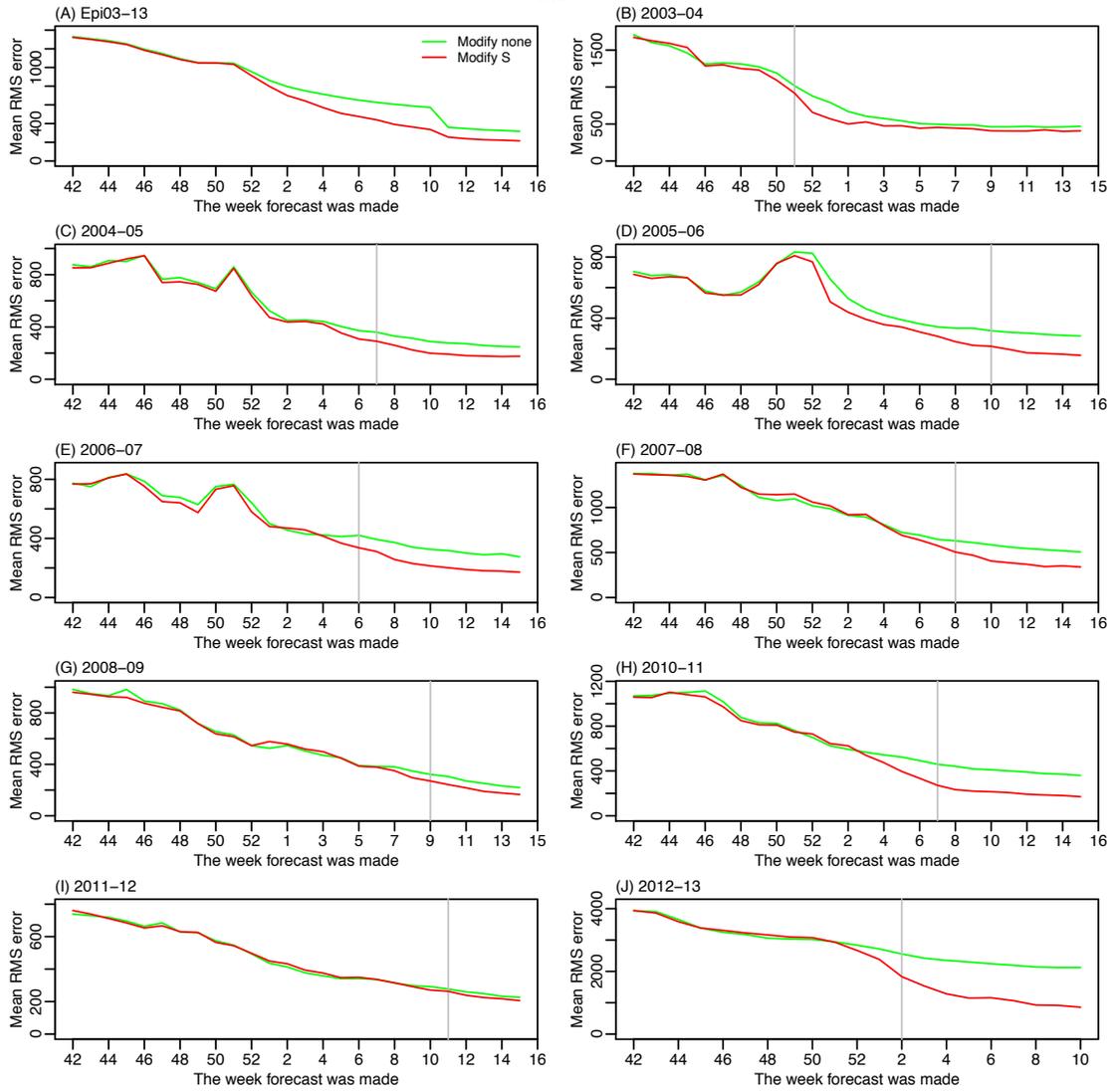

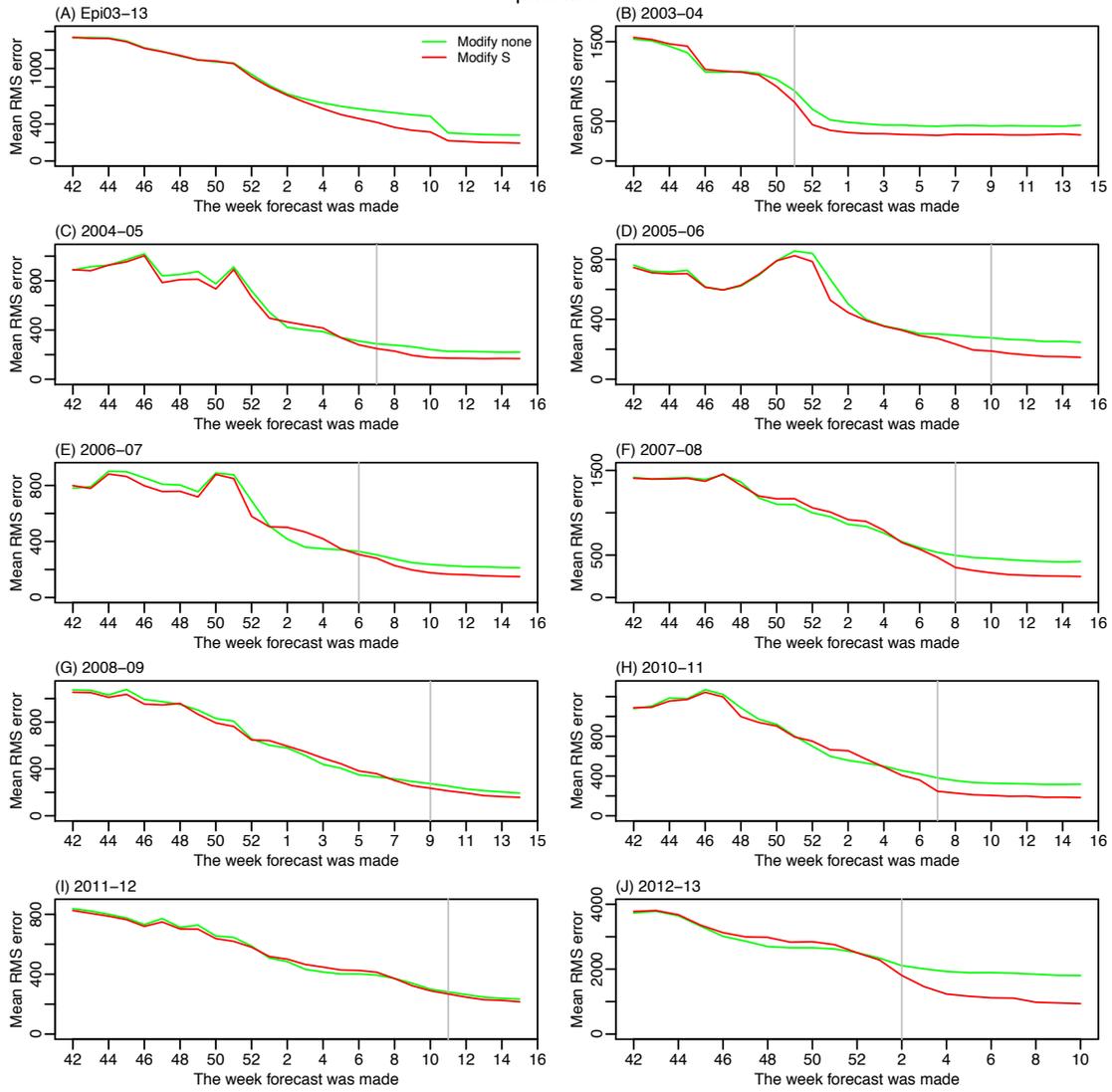

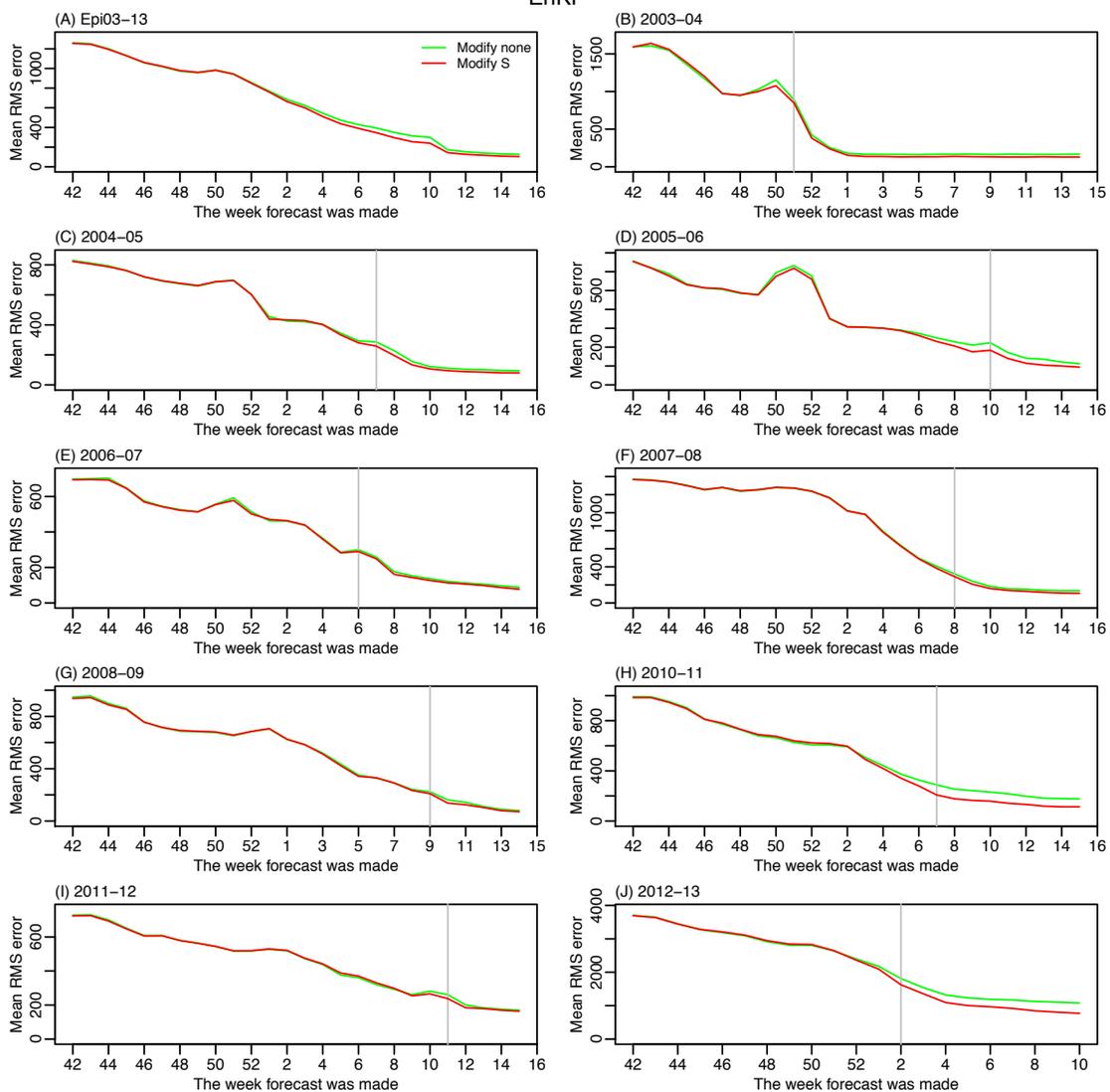

EAKF

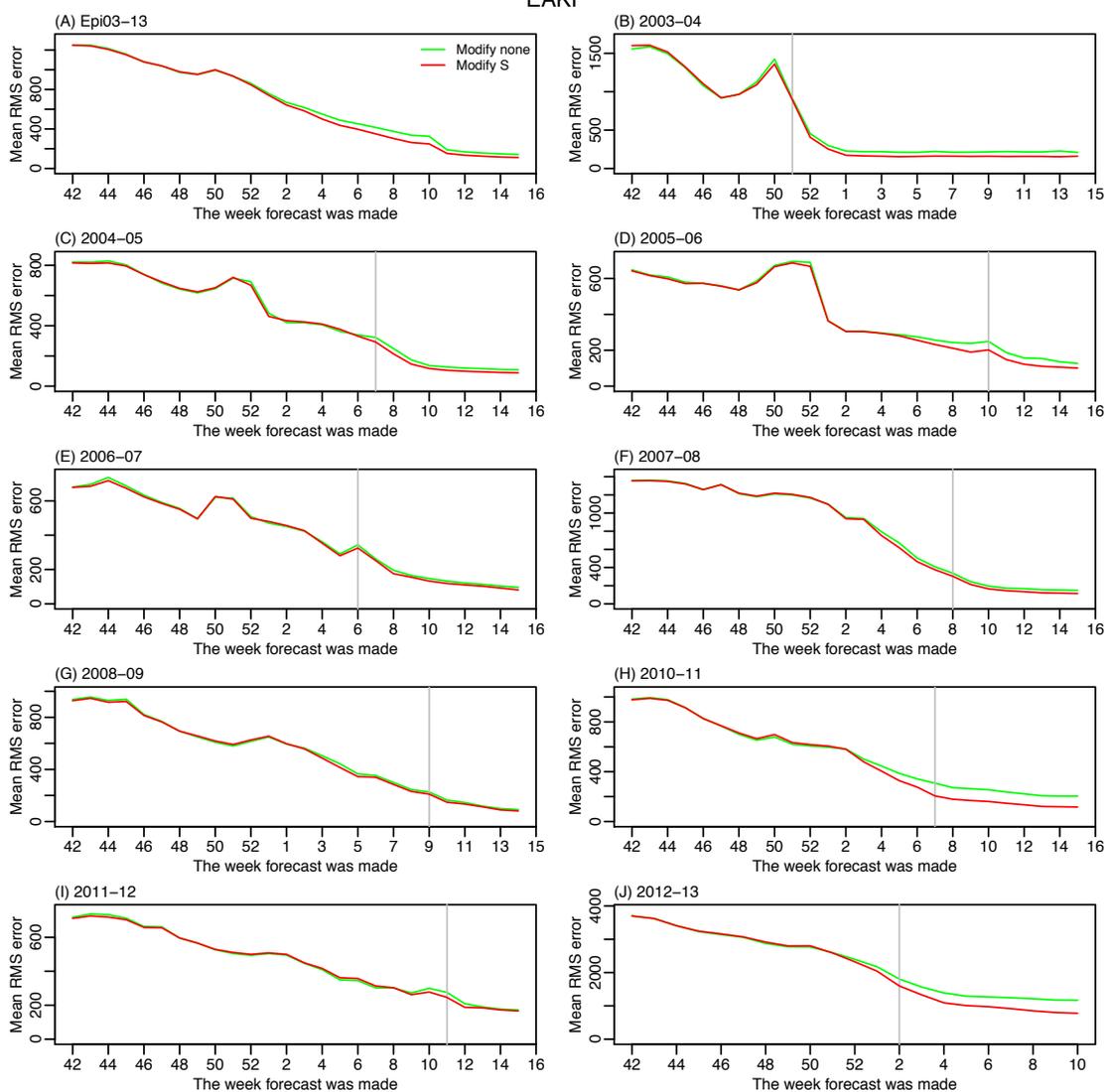

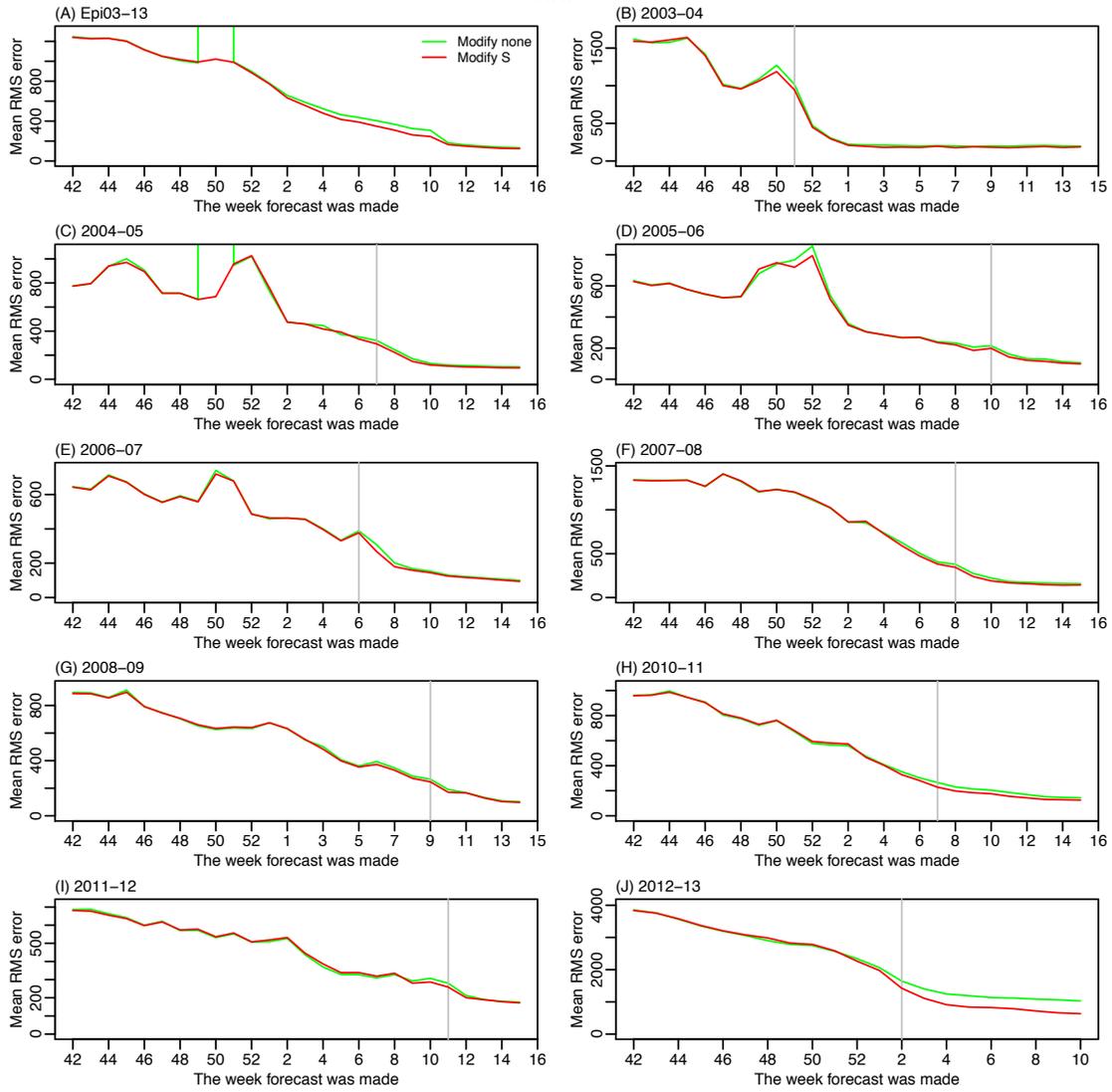